## A table of some coherency matrices, coherency matrix factors (CMFs), and their respective Mueller matrices

**Colin J. R. Sheppard[1,2], Aymeric Le Gratiet[1] and Alberto Diaspro[1,3,4]**

[1] *Nanoscopy and NIC@IIT, Istituto Italiano di Tecnologia, Via Enrico Melen, 83 Edificio B, 16152 Genova, Italy*
[2] *School of Chemistry, University of Wollongong, Wollongong NSW 2522, Australia*
[3] *University of Genoa, 16145 Genoa, Italy*
[4] *Nikon Imaging Center, Istituto Italiano di Tecnologia, via Morego 30, 16163 Genova, Italy*

Many books on polarization give tables of Mueller matrices. Here we give a table of Mueller matrices $\mathbf{M}$, coherency matrices $\mathbf{C}$, and coherency matrix factors $\mathbf{F}$ for different polarization components. $\mathbf{F}$ is not given for some complicated cases. In many cases, though, $\mathbf{F}$ has a very simple form. The terms are defined in the Appendix.

#### 1. Free space

$$\mathbf{M} = \begin{pmatrix} 1 & 0 & 0 & 0 \\ 0 & 1 & 0 & 0 \\ 0 & 0 & 1 & 0 \\ 0 & 0 & 0 & 1 \end{pmatrix}, \mathbf{C} = 2\begin{pmatrix} 1 & 0 & 0 & 0 \\ 0 & 0 & 0 & 0 \\ 0 & 0 & 0 & 0 \\ 0 & 0 & 0 & 0 \end{pmatrix}, \mathbf{F} = \sqrt{2}\begin{pmatrix} 1 & 0 & 0 & 0 \\ 0 & 0 & 0 & 0 \\ 0 & 0 & 0 & 0 \\ 0 & 0 & 0 & 0 \end{pmatrix}.$$

#### 2. Diattenuating retarder

Phase shift $\Delta$.

Diattenuation $D = \cos\kappa = \dfrac{p_1^2 - p_2^2}{p_1^2 + p_2^2}$,

$p_1, p_2$ are the maximum and minimum amplitude coefficients.

Oriented at an angle $\phi$. Terminology as in [1]

$$\mathbf{M} = M_{00}\begin{pmatrix} 1 & \cos\kappa\cos 2\phi & \cos\kappa\sin 2\phi & 0 \\ \cos\kappa\cos 2\phi & \cos^2\phi + \cos\Delta\sin\kappa\sin^2\phi & 0 & -\sin\Delta\sin\kappa\sin 2\phi \\ \sin 2\phi & 0 & \sin^2\phi + \cos\Delta\sin\kappa\cos^2\phi & \sin\Delta\sin\kappa\cos 2\phi \\ 0 & \sin\Delta\sin\kappa\sin 2\phi & -\sin\Delta\sin\kappa\cos 2\phi & \cos\Delta\sin\kappa \end{pmatrix},$$

$$\mathbf{C} = M_{00}\begin{pmatrix} 1+\cos\Delta\sin\kappa & (\cos\kappa - i\sin\Delta\sin\kappa)\cos 2\phi & (\cos\kappa - i\sin\Delta\sin\kappa)\sin 2\phi & 0 \\ (\cos\kappa + i\sin\Delta\sin\kappa)\cos 2\phi & (1-\cos\Delta\sin\kappa)\cos^2 2\phi & (1-\cos\Delta\sin\kappa)\sin 2\phi\cos 2\phi & 0 \\ (\cos\kappa + i\sin\Delta\sin\kappa)\sin 2\phi & (1-\cos\Delta\sin\kappa)\sin 2\phi\cos 2\phi & (1-\cos\Delta\sin\kappa)\sin^2 2\phi & 0 \\ 0 & 0 & 0 & 0 \end{pmatrix},$$

$$\mathbf{F} = \frac{\sqrt{M_{00}}}{\sqrt{1+\cos\Delta\sin\kappa}}\begin{pmatrix} 1+\cos\Delta\sin\kappa & 0 & 0 & 0 \\ (\cos\kappa + i\sin\Delta\sin\kappa)\cos 2\phi & 0 & 0 & 0 \\ (\cos\kappa + i\sin\Delta\sin\kappa)\sin 2\phi & 0 & 0 & 0 \\ 0 & 0 & 0 & 0 \end{pmatrix}.$$

**3. Diattenuating retarder at $0^{\circ}$ (horizontal)**

Phase shift $\Delta$.

Diattenuation $D = \cos\kappa = \dfrac{p_1^2 - p_2^2}{p_1^2 + p_2^2}$,

$p_1, p_2$ are the maximum and minimum amplitude coefficients. Terminology as in [1]

$$\mathbf{M} = M_{00}\begin{pmatrix} 1 & \cos\kappa & 0 & 0 \\ \cos\kappa & 1 & 0 & 0 \\ 0 & 0 & \cos\Delta\sin\kappa & \sin\Delta\sin\kappa \\ 0 & 0 & -\sin\Delta\sin\kappa & \cos\Delta\sin\kappa \end{pmatrix}, \mathbf{C} = M_{00}\begin{pmatrix} 1+\cos\Delta\sin\kappa & \cos\kappa - i\sin\Delta\sin\kappa & 0 & 0 \\ \cos\kappa + i\sin\Delta\sin\kappa & 1-\cos\Delta\sin\kappa & 0 & 0 \\ 0 & 0 & 0 & 0 \\ 0 & 0 & 0 & 0 \end{pmatrix},$$

$$\mathbf{F} = \frac{\sqrt{M_{00}}}{\sqrt{1+\cos\Delta\sin\kappa}}\begin{pmatrix} 1+\cos\Delta\sin\kappa & 0 & 0 & 0 \\ \cos\kappa + i\sin\Delta\sin\kappa & 0 & 0 & 0 \\ 0 & 0 & 0 & 0 \\ 0 & 0 & 0 & 0 \end{pmatrix}.$$

**4. Elliptic diattenuator**

$$D = \cos\kappa = \frac{p_1^2 - p_2^2}{p_1^2 + p_2^2}$$

The ratio between the strengths of components of the electric field = tan $\chi$, ellipticity angle = $\chi$.

Oriented at an angle $\phi$.

Terminology as in [1].

$$\mathbf{M} = \frac{1}{4}\begin{pmatrix} 2(p_1^2+p_2^2) & 2(p_1^2-p_2^2)\cos 2\chi\cos 2\phi & 2(p_1^2-p_2^2)\cos 2\chi\sin 2\phi & 2(p_1^2-p_2^2)\sin 2\chi \\ 2(p_1^2-p_2^2)\cos 2\chi\cos 2\phi & p_2(3p_1+p_2)+2(p_1-p_2)^2\cos^2 2\chi\cos^2 2\phi & (p_1-p_2)^2\cos^2 2\chi\sin 4\phi & (p_1-p_2)^2\sin 4\chi\cos 2\phi \\ 2(p_1^2-p_2^2)\cos 2\chi\sin 2\phi & (p_1-p_2)^2\cos^2 2\chi\sin 4\phi & p_2(3p_1+p_2)+2(p_1-p_2)^2\cos^2 2\chi\sin^2 2\phi & (p_1-p_2)^2\sin 4\chi\sin 2\phi \\ 2(p_1^2-p_2^2)\sin 2\chi & (p_1-p_2)^2\sin 4\chi\cos 2\phi & (p_1-p_2)^2\sin 4\chi\sin 2\phi & (p_1+p_2)^2+(p_1-p_2)^2\cos 4\chi \end{pmatrix},$$

$$\mathbf{C} = \frac{1}{2}\begin{pmatrix} (p_1+p_2)^2 & (p_1^2-p_2^2)\cos 2\chi\cos 2\phi & (p_1^2-p_2^2)\cos 2\chi\sin 2\phi & (p_1^2-p_2^2)\sin 2\chi \\ (p_1^2-p_2^2)\cos 2\chi\cos 2\phi & (p_1-p_2)^2\cos^2 2\chi\cos^2 2\phi & (p_1^2-p_2^2)\cos^2 2\chi\sin 2\phi\cos 2\phi & (p_1^2-p_2^2)\sin 2\chi\cos 2\chi\cos 2\phi \\ (p_1^2-p_2^2)\cos 2\chi\sin 2\phi & (p_1^2-p_2^2)\cos^2 2\chi\sin 2\phi\cos 2\phi & (p_1-p_2)^2\cos^2 2\chi\sin^2 2\phi & (p_1^2-p_2^2)\sin 2\chi\cos 2\chi\sin 2\phi \\ (p_1^2-p_2^2)\sin 2\chi & (p_1^2-p_2^2)\sin 2\chi\cos 2\chi\cos 2\phi & (p_1^2-p_2^2)\sin 2\chi\cos 2\chi\sin 2\phi & (p_1-p_2)^2\sin^2 2\chi \end{pmatrix},$$

$$\mathbf{F} = \frac{1}{\sqrt{2}}\begin{pmatrix} p_1+p_2 & 0 & 0 & 0 \\ (p_1-p_2)\cos 2\chi\cos 2\phi & 0 & 0 & 0 \\ (p_1-p_2)\cos 2\chi\sin 2\phi & 0 & 0 & 0 \\ (p_1-p_2)\sin 2\chi & 0 & 0 & 0 \end{pmatrix}.$$

**5. Elliptic diattenuator at $0^{\circ}$**

$$D = \cos\kappa = \frac{p_1^2 - p_2^2}{p_1^2 + p_2^2}$$

The ratio between the strengths of components of the electric field = tan $\chi$, ellipticity angle = $\chi$.

Terminology as in [1].

$$\mathbf{M} = \frac{p_1^2}{1+\cos\kappa}\begin{pmatrix} 1 & \cos\kappa\cos 2\chi & 0 & \cos\kappa\sin 2\chi \\ \cos\kappa\cos 2\chi & \sin\kappa + (1-\sin\kappa)\cos^2 2\chi & 0 & (1-\sin\kappa)\sin 2\chi\cos 2\chi \\ 0 & 0 & \sin\kappa & 0 \\ \cos\kappa\sin 2\chi & (1-\sin\kappa)\sin 2\chi\cos 2\chi & 0 & \sin\kappa + (1-\sin\kappa)\sin^2 2\chi \end{pmatrix},$$

$$\mathbf{C} = \frac{p_1^2}{1+\cos\kappa}\begin{pmatrix} 1+\sin\kappa & \cos\kappa\cos 2\chi & 0 & \cos\kappa\sin 2\chi \\ \cos\kappa\cos 2\chi & (1-\sin\kappa)\cos^2 2\chi & 0 & (1-\sin\kappa)\sin 2\chi\cos 2\chi \\ 0 & 0 & 0 & 0 \\ \cos\kappa\sin 2\chi & (1-\sin\kappa)\sin 2\chi\cos 2\chi & 0 & (1-\sin\kappa)\sin^2 2\chi \end{pmatrix},$$

$$\mathbf{F}=\frac{p_1}{\sqrt{1+\cos\kappa}}\cdot\begin{pmatrix}\sqrt{1+\sin\kappa} & 0 & 0 & 0\\ \sqrt{1-\sin\kappa}\cos 2\chi & 0 & 0 & 0\\ 0 & 0 & 0 & 0\\ \sqrt{1-\sin\kappa}\sin 2\chi & 0 & 0 & 0\end{pmatrix}=\frac{1}{\sqrt{2}}\begin{pmatrix}p_1+p_2 & 0 & 0 & 0\\ (p_1-p_2)\cos 2\chi & 0 & 0 & 0\\ 0 & 0 & 0 & 0\\ (p_1-p_2)\sin 2\chi & 0 & 0 & 0\end{pmatrix}.$$

**6. Elliptic diattenuator at** $45^{\circ}$

$$\mathbf{M}=\frac{1}{4}\begin{pmatrix}2(p_1^2+p_2^2) & 0 & 2(p_1^2-p_2^2)\cos 2\chi & 2(p_1^2-p_2^2)\sin 2\chi\\ 0 & p_2(3p_1+p_2) & 0 & 0\\ 2(p_1^2-p_2^2)\cos 2\chi & 0 & p_2(3p_1+p_2)+2(p_1-p_2)^2\cos^2 2\chi & (p_1-p_2)^2\sin 4\chi\\ 2(p_1^2-p_2^2)\sin 2\chi & 0 & (p_1-p_2)^2\sin 4\chi & (p_1+p_2)^2+(p_1-p_2)^2\cos 4\chi\end{pmatrix},$$

$$\mathbf{C}=\frac{1}{2}\begin{pmatrix}(p_1+p_2)^2 & 0 & (p_1^2-p_2^2)\cos 2\chi & (p_1^2-p_2^2)\sin 2\chi\\ 0 & 0 & 0 & 0\\ (p_1^2-p_2^2)\cos 2\chi & 0 & (p_1-p_2)^2\cos^2 2\chi & (p_1^2-p_2^2)\sin 2\chi\cos 2\chi\\ (p_1^2-p_2^2)\sin 2\chi & 0 & (p_1^2-p_2^2)\sin 2\chi\cos 2\chi & (p_1-p_2)^2\sin^2 2\chi\end{pmatrix},$$

$$\mathbf{F}=\frac{1}{\sqrt{2}}\begin{pmatrix}p_1+p_2 & 0 & 0 & 0\\ 0 & 0 & 0 & 0\\ (p_1-p_2)\cos 2\chi & 0 & 0 & 0\\ (p_1-p_2)\sin 2\chi & 0 & 0 & 0\end{pmatrix}.$$

**7. Elliptic diattenuator at** $90^{\circ}$

$$\mathbf{M}=\frac{1}{4}\begin{pmatrix}2(p_1^2+p_2^2) & -2(p_1^2-p_2^2)\cos 2\chi & 0 & 2(p_1^2-p_2^2)\sin 2\chi\\ -2(p_1^2-p_2^2)\cos 2\chi & p_2(3p_1+p_2)+2(p_1-p_2)^2\cos^2 2\chi & 0 & -(p_1-p_2)^2\sin 4\chi\\ 0 & 0 & p_2(3p_1+p_2) & 0\\ 2(p_1^2-p_2^2)\sin 2\chi & -(p_1-p_2)^2\sin 4\chi & 0 & (p_1+p_2)^2+(p_1-p_2)^2\cos 4\chi\end{pmatrix},$$

$$\mathbf{C}=\frac{1}{2}\begin{pmatrix}(p_1+p_2)^2 & -(p_1^2-p_2^2)\cos 2\chi & 0 & (p_1^2-p_2^2)\sin 2\chi\\ -(p_1^2-p_2^2)\cos 2\chi & (p_1-p_2)^2\cos^2 2\chi & 0 & -(p_1^2-p_2^2)\sin 2\chi\cos 2\chi\\ 0 & 0 & 0 & 0\\ (p_1^2-p_2^2)\sin 2\chi & -(p_1^2-p_2^2)\sin 2\chi\cos 2\chi & 0 & (p_1-p_2)^2\sin^2 2\chi\end{pmatrix},$$

$$\mathbf{F}=\frac{1}{\sqrt{2}}\begin{pmatrix}p_1+p_2 & 0 & 0 & 0\\ -(p_1-p_2)\cos 2\chi & 0 & 0 & 0\\ 0 & 0 & 0 & 0\\ (p_1-p_2)\sin 2\chi & 0 & 0 & 0\end{pmatrix}.$$

**8. Elliptic diattenuator at** $135^\circ$

$$\mathbf{M}=\frac{1}{4}\begin{pmatrix} 2(p_1^2+p_2^2) & 0 & -2(p_1^2-p_2^2)\cos 2\chi & 2(p_1^2-p_2^2)\sin 2\chi \\ 0 & p_2(3p_1+p_2) & 0 & 0 \\ -2(p_1^2-p_2^2)\cos 2\chi & 0 & p_2(3p_1+p_2)+2(p_1-p_2)^2\cos^2 2\chi & -(p_1-p_2)^2\sin 4\chi \\ 2(p_1^2-p_2^2)\sin 2\chi & 0 & -(p_1-p_2)^2\sin 4\chi & (p_1+p_2)^2+(p_1-p_2)^2\cos 4\chi \end{pmatrix},$$

$$\mathbf{C}=\frac{1}{2}\begin{pmatrix} (p_1+p_2)^2 & 0 & -(p_1^2-p_2^2)\cos 2\chi & (p_1^2-p_2^2)\sin 2\chi \\ 0 & 0 & 0 & 0 \\ -(p_1^2-p_2^2)\cos 2\chi & 0 & (p_1-p_2)^2\cos^2 2\chi & -(p_1^2-p_2^2)\sin 2\chi\cos 2\chi \\ (p_1^2-p_2^2)\sin 2\chi & 0 & -(p_1^2-p_2^2)\sin 2\chi\cos 2\chi & (p_1-p_2)^2\sin^2 2\chi \end{pmatrix},$$

$$\mathbf{F}=\frac{1}{\sqrt{2}}\begin{pmatrix} p_1+p_2 & 0 & 0 & 0 \\ 0 & 0 & 0 & 0 \\ -(p_1-p_2)\cos 2\chi & 0 & 0 & 0 \\ (p_1-p_2)\sin 2\chi & 0 & 0 & 0 \end{pmatrix}.$$

**9. Linear diattenuator**

Oriented at an angle $\phi$.

$D=\cos\kappa=\dfrac{p_1^2-p_2^2}{p_1^2+p_2^2}$.

Terminology as in [1].

$$\mathbf{M}=\frac{p_1^2}{1+\cos\kappa}\begin{pmatrix} 1 & \cos\kappa\cos 2\phi & \cos\kappa\sin 2\phi & 0 \\ \cos\kappa\cos 2\phi & \cos^2 2\phi+\sin\kappa\sin^2 2\phi & (1-\sin\kappa)\sin 2\phi\cos 2\phi & 0 \\ \cos\kappa\sin 2\phi & (1-\sin\kappa)\sin 2\phi\cos 2\phi & \sin^2 2\phi+\sin\kappa\cos^2 2\phi & 0 \\ 0 & 0 & 0 & \sin\kappa \end{pmatrix},$$

$$\mathbf{C}=\frac{p_1^2}{1+\cos\kappa}\begin{pmatrix} 1+\sin\kappa & \cos\kappa\cos 2\phi & \cos\kappa\sin 2\phi & 0 \\ \cos\kappa\cos 2\phi & (1-\sin\kappa)\cos^2 2\phi & (1-\sin\kappa)\sin 2\phi\cos 2\phi & 0 \\ \cos\kappa\sin 2\phi & (1-\sin\kappa)\sin 2\phi\cos 2\phi & (1-\sin\kappa)\sin^2 2\phi & 0 \\ 0 & 0 & 0 & 0 \end{pmatrix},$$

$$\mathbf{F}=\frac{p_1}{\sqrt{1+\cos\kappa}}\begin{pmatrix} \sqrt{1+\sin\kappa} & 0 & 0 & 0 \\ \sqrt{1-\sin\kappa}\cos 2\phi & 0 & 0 & 0 \\ \sqrt{1-\sin\kappa}\sin 2\phi & 0 & 0 & 0 \\ 0 & 0 & 0 & 0 \end{pmatrix}=\frac{1}{\sqrt{2}}\begin{pmatrix} p_1+p_2 & 0 & 0 & 0 \\ (p_1-p_2)\cos 2\phi & 0 & 0 & 0 \\ (p_1-p_2)\sin 2\phi & 0 & 0 & 0 \\ 0 & 0 & 0 & 0 \end{pmatrix}.$$

**10. Linear diattenuator at** $0^\circ$

$\chi = 0$ , $M_{00} = \frac{1}{2}\left(p_1^2 + p_2^2\right) = \dfrac{p_1^2}{1+\cos\kappa}$ . Terminology as in [1].

$$\mathbf{M} = \frac{1}{2}\begin{pmatrix} p_1^2+p_2^2 & p_1^2-p_2^2 & 0 & 0 \\ p_1^2-p_2^2 & p_1^2+p_2^2 & 0 & 0 \\ 0 & 0 & 2p_1p_2 & 0 \\ 0 & 0 & 0 & 2p_1p_2 \end{pmatrix} = \frac{p_1^2}{1+\cos\kappa}\begin{pmatrix} 1 & \cos\kappa & 0 & 0 \\ \cos\kappa & 1 & 0 & 0 \\ 0 & 0 & \sin\kappa & 0 \\ 0 & 0 & 0 & \sin\kappa \end{pmatrix},$$

$$\mathbf{C} = \frac{1}{2}\begin{pmatrix} (p_1+p_2)^2 & p_1^2-p_2^2 & 0 & 0 \\ p_1^2-p_2^2 & (p_1-p_2)^2 & 0 & 0 \\ 0 & 0 & 0 & 0 \\ 0 & 0 & 0 & 0 \end{pmatrix} = \frac{p_1^2}{1+\cos\kappa}\begin{pmatrix} 1+\sin\kappa & \cos\kappa & 0 & 0 \\ \cos\kappa & 1-\sin\kappa & 0 & 0 \\ 0 & 0 & 0 & 0 \\ 0 & 0 & 0 & 0 \end{pmatrix},$$

$$\mathbf{F} = \frac{1}{\sqrt{2}}\begin{pmatrix} p_1+p_2 & 0 & 0 & 0 \\ p_1-p_2 & 0 & 0 & 0 \\ 0 & 0 & 0 & 0 \\ 0 & 0 & 0 & 0 \end{pmatrix} = \frac{p_1}{\sqrt{1+\cos\kappa}}\begin{pmatrix} \sqrt{1+\sin\kappa} & 0 & 0 & 0 \\ \sqrt{1-\sin\kappa} & 0 & 0 & 0 \\ 0 & 0 & 0 & 0 \\ 0 & 0 & 0 & 0 \end{pmatrix}.$$

**11. Linear diattenuator at** $45^\circ$

$$\mathbf{M} = \frac{p_1^2}{1+\cos\kappa}\begin{pmatrix} 1 & 0 & \cos\kappa & 0 \\ 0 & \sin\kappa & 0 & 0 \\ \cos\kappa & 0 & 1 & 0 \\ 0 & 0 & 0 & \sin\kappa \end{pmatrix},$$

$$\mathbf{C} = \frac{p_1^2}{1+\cos\kappa}\begin{pmatrix} 1+\sin\kappa & 0 & \cos\kappa & 0 \\ 0 & 0 & 0 & 0 \\ \cos\kappa & 0 & 1-\sin\kappa & 0 \\ 0 & 0 & 0 & 0 \end{pmatrix},$$

$$\mathbf{F} = \frac{p_1}{\sqrt{1+\cos\kappa}}\begin{pmatrix} \sqrt{1+\sin\kappa} & 0 & 0 & 0 \\ 0 & 0 & 0 & 0 \\ \sqrt{1-\sin\kappa} & 0 & 0 & 0 \\ 0 & 0 & 0 & 0 \end{pmatrix} = \frac{1}{\sqrt{2}}\begin{pmatrix} p_1+p_2 & 0 & 0 & 0 \\ 0 & 0 & 0 & 0 \\ p_1-p_2 & 0 & 0 & 0 \\ 0 & 0 & 0 & 0 \end{pmatrix}.$$

**12. Linear diattenuator at** $90^\circ$

$$\mathbf{M} = \frac{p_1^2}{1+\cos\kappa}\begin{pmatrix} 1 & -\cos\kappa & 0 & 0 \\ -\cos\kappa & 1 & 0 & 0 \\ 0 & 0 & \sin\kappa & 0 \\ 0 & 0 & 0 & \sin\kappa \end{pmatrix},$$

$$\mathbf{C} = \frac{p_1^2}{1+\cos\kappa}\begin{pmatrix} 1+\sin\kappa & -\cos\kappa & 0 & 0 \\ -\cos\kappa & 1-\sin\kappa & 0 & 0 \\ 0 & 0 & 0 & 0 \\ 0 & 0 & 0 & 0 \end{pmatrix},$$

$$\mathbf{F} = \frac{p_1}{\sqrt{1+\cos\kappa}}\begin{pmatrix} \sqrt{1+\sin\kappa} & 0 & 0 & 0 \\ -\sqrt{1-\sin\kappa} & 0 & 0 & 0 \\ 0 & 0 & 0 & 0 \\ 0 & 0 & 0 & 0 \end{pmatrix} = \frac{1}{\sqrt{2}}\begin{pmatrix} p_1+p_2 & 0 & 0 & 0 \\ -(p_1-p_2) & 0 & 0 & 0 \\ 0 & 0 & 0 & 0 \\ 0 & 0 & 0 & 0 \end{pmatrix}.$$

**13. Linear diattenuator at $135^{\circ}$**

$$\mathbf{M}=\frac{p_1^2}{1+\cos\kappa}\begin{pmatrix} 1 & 0 & -\cos\kappa & 0 \\ 0 & \sin\kappa & 0 & 0 \\ -\cos\kappa & 0 & 1 & 0 \\ 0 & 0 & 0 & \sin\kappa \end{pmatrix},$$

$$\mathbf{C}=\frac{p_1^2}{1+\cos\kappa}\begin{pmatrix} 1+\sin\kappa & 0 & -\cos\kappa & 0 \\ 0 & 0 & 0 & 0 \\ -\cos\kappa & 0 & (1-\sin\kappa) & 0 \\ 0 & 0 & 0 & 0 \end{pmatrix},$$

$$\mathbf{F}=\frac{p_1}{\sqrt{1+\cos\kappa}}\begin{pmatrix} \sqrt{1+\sin\kappa} & 0 & 0 & 0 \\ 0 & 0 & 0 & 0 \\ -\sqrt{1-\sin\kappa} & 0 & 0 & 0 \\ 0 & 0 & 0 & 0 \end{pmatrix}=\frac{1}{\sqrt{2}}\begin{pmatrix} p_1+p_2 & 0 & 0 & 0 \\ 0 & 0 & 0 & 0 \\ -(p_1-p_2) & 0 & 0 & 0 \\ 0 & 0 & 0 & 0 \end{pmatrix}.$$

**14. Elliptic polarizer at $0^{\circ}$**

The ratio between the strengths of components of the electric field = tan $\chi$ , ellipticity angle = $\chi$ . Terminology as in [1].

$$\mathbf{M}=\frac{p_1^2}{2}\begin{pmatrix} 1 & \cos 2\chi & 0 & \sin 2\chi \\ \cos 2\chi & \cos^2 2\chi & 0 & \sin 2\chi\cos 2\chi \\ 0 & 0 & 0 & 0 \\ \sin 2\chi & \sin 2\chi\cos 2\chi & 0 & \sin^2 2\chi \end{pmatrix},\mathbf{C}=\frac{p_1^2}{2}\begin{pmatrix} 1 & \cos 2\chi & 0 & \sin 2\chi \\ \cos 2\chi & \cos^2 2\chi & 0 & \sin 2\chi\cos 2\chi \\ 0 & 0 & 0 & 0 \\ \sin 2\chi & \sin 2\chi\cos 2\chi & 0 & \sin^2 2\chi \end{pmatrix},$$

$$\mathbf{F}=\frac{p_1}{\sqrt{2}}\begin{pmatrix} 1 & 0 & 0 & 0 \\ \cos 2\chi & 0 & 0 & 0 \\ 0 & 0 & 0 & 0 \\ \sin 2\chi & 0 & 0 & 0 \end{pmatrix}.$$

**15. Elliptic polarizer at $45^{\circ}$**

$$\mathbf{M}=\frac{p_1^2}{2}\begin{pmatrix} 1 & 0 & \cos 2\chi & \sin 2\chi \\ 0 & 0 & 0 & 0 \\ \cos 2\chi & 0 & \cos^2 2\chi & \sin 2\chi\cos 2\chi \\ \sin 2\chi & 0 & \sin 2\chi\cos 2\chi & \cos^2 2\chi \end{pmatrix},$$

$$\mathbf{C}=\frac{p_1^2}{2}\begin{pmatrix} 1 & 0 & \cos 2\chi & \sin 2\chi \\ 0 & 0 & 0 & 0 \\ \cos 2\chi & 0 & \cos^2 2\chi & \sin 2\chi\cos 2\chi \\ \sin 2\chi & 0 & \sin 2\chi\cos 2\chi & \sin^2 2\chi \end{pmatrix},$$

$$\mathbf{F}=\frac{p_1}{\sqrt{2}}\begin{pmatrix} 1 & 0 & 0 & 0 \\ 0 & 0 & 0 & 0 \\ \cos 2\chi & 0 & 0 & 0 \\ \sin 2\chi & 0 & 0 & 0 \end{pmatrix}.$$

**16. Elliptic polarizer at $90^{\circ}$**

$$\mathbf{M}=\frac{p_1^2}{2}\begin{pmatrix} 1 & -\cos 2\chi & 0 & \sin 2\chi \\ -\cos 2\chi & \cos^2 2\chi & 0 & -\sin 2\chi \cos 2\chi \\ 0 & 0 & 0 & 0 \\ \sin 2\chi & -\sin 2\chi \cos 2\chi & 0 & \cos^2 2\chi \end{pmatrix},$$

$$\mathbf{C}=\frac{p_1^2}{2}\begin{pmatrix} 1 & -\cos 2\chi & 0 & \sin 2\chi \\ -\cos 2\chi & \cos^2 2\chi & 0 & -\sin 2\chi \cos 2\chi \\ 0 & 0 & 0 & 0 \\ \sin 2\chi & -\sin 2\chi \cos 2\chi & 0 & \sin^2 2\chi \end{pmatrix},$$

$$\mathbf{F}=\frac{p_1}{\sqrt{2}}\begin{pmatrix} 1 & 0 & 0 & 0 \\ -\cos 2\chi & 0 & 0 & 0 \\ 0 & 0 & 0 & 0 \\ \sin 2\chi & 0 & 0 & 0 \end{pmatrix}.$$

**17. Elliptic polarizer at $135^{\circ}$**

$$\mathbf{M}=\frac{p_1^2}{2}\begin{pmatrix} 1 & 0 & -\cos 2\chi & \sin 2\chi \\ 0 & 0 & 0 & 0 \\ -\cos 2\chi & 0 & \cos^2 2\chi & -\sin 2\chi \cos 2\chi \\ \sin 2\chi & 0 & -\sin 2\chi \cos 2\chi & \cos^2 2\chi \end{pmatrix},$$

$$\mathbf{C}=\frac{p_1^2}{2}\begin{pmatrix} 1 & 0 & -\cos 2\chi & \sin 2\chi \\ 0 & 0 & 0 & 0 \\ -\cos 2\chi & 0 & \cos^2 2\chi & -\sin 2\chi \cos 2\chi \\ \sin 2\chi & 0 & -\sin 2\chi \cos 2\chi & \sin^2 2\chi \end{pmatrix},$$

$$\mathbf{F}=\frac{p_1}{\sqrt{2}}\begin{pmatrix} 1 & 0 & 0 & 0 \\ 0 & 0 & 0 & 0 \\ -\cos 2\chi & 0 & 0 & 0 \\ \sin 2\chi & 0 & 0 & 0 \end{pmatrix}.$$

**18. Linear total polarizer**

Oriented at an angle $\phi$. Terminology as in [1].

$$\mathbf{M}=\frac{p_1^2}{2}\begin{pmatrix} 1 & \cos 2\phi & \sin 2\phi & 0 \\ \cos 2\phi & \cos^2 2\phi & \sin 2\phi \cos 2\phi & 0 \\ \sin 2\phi & \sin 2\phi \cos 2\phi & \sin^2 2\phi & 0 \\ 0 & 0 & 0 & 0 \end{pmatrix}, \mathbf{C}=\frac{p_1^2}{2}\begin{pmatrix} 1 & \cos 2\phi & \sin 2\phi & 0 \\ \cos 2\phi & \cos^2 2\phi & \sin 2\phi \cos 2\phi & 0 \\ \sin 2\phi & \sin 2\phi \cos 2\phi & \sin^2 2\phi & 0 \\ 0 & 0 & 0 & 0 \end{pmatrix},$$

$$\mathbf{F}=\frac{p_1}{\sqrt{2}}\begin{pmatrix} 1 & 0 & 0 & 0 \\ \cos 2\phi & 0 & 0 & 0 \\ \sin 2\phi & 0 & 0 & 0 \\ 0 & 0 & 0 & 0 \end{pmatrix}.$$

**19. Linear polarizer at** $0^\circ$

$$\mathbf{M}=\frac{1}{2}\begin{pmatrix} 1 & 1 & 0 & 0 \\ 1 & 1 & 0 & 0 \\ 0 & 0 & 0 & 0 \\ 0 & 0 & 0 & 0 \end{pmatrix},\mathbf{C}=\frac{1}{2}\begin{pmatrix} 1 & 1 & 0 & 0 \\ 1 & 1 & 0 & 0 \\ 0 & 0 & 0 & 0 \\ 0 & 0 & 0 & 0 \end{pmatrix},\mathbf{F}=\frac{1}{\sqrt{2}}\begin{pmatrix} 1 & 0 & 0 & 0 \\ 1 & 0 & 0 & 0 \\ 0 & 0 & 0 & 0 \\ 0 & 0 & 0 & 0 \end{pmatrix}.$$

**20. Linear polarizer at** $45^\circ$

$$\mathbf{M}=\frac{1}{2}\begin{pmatrix} 1 & 0 & 1 & 0 \\ 0 & 0 & 0 & 0 \\ 1 & 0 & 1 & 0 \\ 0 & 0 & 0 & 0 \end{pmatrix},\mathbf{C}=\frac{1}{2}\begin{pmatrix} 1 & 0 & 1 & 0 \\ 0 & 0 & 0 & 0 \\ 1 & 0 & 1 & 0 \\ 0 & 0 & 0 & 0 \end{pmatrix},\mathbf{F}=\frac{1}{\sqrt{2}}\begin{pmatrix} 1 & 0 & 0 & 0 \\ 0 & 0 & 0 & 0 \\ 1 & 0 & 0 & 0 \\ 0 & 0 & 0 & 0 \end{pmatrix}.$$

**21. Linear polarizer at** $90^\circ$

$$\mathbf{M}=\frac{1}{2}\begin{pmatrix} 1 & -1 & 0 & 0 \\ -1 & 1 & 0 & 0 \\ 0 & 0 & 0 & 0 \\ 0 & 0 & 0 & 0 \end{pmatrix},\mathbf{C}=\frac{1}{2}\begin{pmatrix} 1 & -1 & 0 & 0 \\ -1 & 1 & 0 & 0 \\ 0 & 0 & 0 & 0 \\ 0 & 0 & 0 & 0 \end{pmatrix},\mathbf{F}=\frac{1}{\sqrt{2}}\begin{pmatrix} 1 & 0 & 0 & 0 \\ -1 & 0 & 0 & 0 \\ 0 & 0 & 0 & 0 \\ 0 & 0 & 0 & 0 \end{pmatrix}.$$

**22. Linear polarizer at** $135^\circ$

$$\mathbf{M}=\frac{1}{2}\begin{pmatrix} 1 & 0 & -1 & 0 \\ 0 & 0 & 0 & 0 \\ -1 & 0 & 1 & 0 \\ 0 & 0 & 0 & 0 \end{pmatrix},\mathbf{C}=\frac{1}{2}\begin{pmatrix} 1 & 0 & -1 & 0 \\ 0 & 0 & 0 & 0 \\ -1 & 0 & 1 & 0 \\ 0 & 0 & 0 & 0 \end{pmatrix},\mathbf{F}=\frac{1}{\sqrt{2}}\begin{pmatrix} 1 & 0 & 0 & 0 \\ 0 & 0 & 0 & 0 \\ -1 & 0 & 0 & 0 \\ 0 & 0 & 0 & 0 \end{pmatrix}.$$

**23. Circular diattenuator**

Diattenuation $D=\cos\kappa=\dfrac{p_1^2-p_2^2}{p_1^2+p_2^2}$,

$M_{00}=\tfrac{1}{2}\left(p_1^2+p_2^2\right)=\dfrac{p_1^2}{1+\cos\kappa}$,

$p_1, p_2$ are the maximum and minimum amplitude coefficients. Terminology as in [1].

$$\mathbf{M}=\frac{p_1^2}{1+\cos\kappa}\begin{pmatrix} 1 & 0 & 0 & \cos\kappa \\ 0 & \sin\kappa & 0 & 0 \\ 0 & 0 & \sin\kappa & 0 \\ \cos\kappa & 0 & 0 & 1 \end{pmatrix},\mathbf{C}=\frac{p_1^2}{1+\cos\kappa}\begin{pmatrix} 1+\sin\kappa & 0 & 0 & \cos\kappa \\ 0 & 0 & 0 & 0 \\ 0 & 0 & 0 & 0 \\ \cos\kappa & 0 & 0 & 1-\sin\kappa \end{pmatrix},$$

$$\mathbf{F}=\frac{p_1}{\sqrt{1+\cos\kappa}}\begin{pmatrix} \sqrt{1+\sin\kappa} & 0 & 0 & 0 \\ 0 & 0 & 0 & 0 \\ 0 & 0 & 0 & 0 \\ \sqrt{1-\sin\kappa} & 0 & 0 & 0 \end{pmatrix}=\frac{1}{\sqrt{2}}\begin{pmatrix} p_1+p_2 & 0 & 0 & 0 \\ 0 & 0 & 0 & 0 \\ 0 & 0 & 0 & 0 \\ p_1-p_2 & 0 & 0 & 0 \end{pmatrix}.$$

**24. Circular polarizer, left**

$$\mathbf{M}=\frac{1}{2}\begin{pmatrix} 1 & 0 & 0 & 1 \\ 0 & 0 & 0 & 0 \\ 0 & 0 & 0 & 0 \\ 1 & 0 & 0 & 1 \end{pmatrix},\mathbf{C}=\frac{1}{2}\begin{pmatrix} 1 & 0 & 0 & 1 \\ 0 & 0 & 0 & 0 \\ 0 & 0 & 0 & 0 \\ 1 & 0 & 0 & 1 \end{pmatrix},\mathbf{F}=\frac{1}{\sqrt{2}}\begin{pmatrix} 1 & 0 & 0 & 0 \\ 0 & 0 & 0 & 0 \\ 0 & 0 & 0 & 0 \\ 1 & 0 & 0 & 0 \end{pmatrix}.$$

### 25. Circular polarizer, right

$$\mathbf{M}=\frac{1}{2}\begin{pmatrix} 1 & 0 & 0 & -1 \\ 0 & 0 & 0 & 0 \\ 0 & 0 & 0 & 0 \\ -1 & 0 & 0 & 1 \end{pmatrix},\mathbf{C}=\frac{1}{2}\begin{pmatrix} 1 & 0 & 0 & -1 \\ 0 & 0 & 0 & 0 \\ 0 & 0 & 0 & 0 \\ -1 & 0 & 0 & 1 \end{pmatrix},\mathbf{F}=\frac{1}{\sqrt{2}}\begin{pmatrix} 1 & 0 & 0 & 0 \\ 0 & 0 & 0 & 0 \\ 0 & 0 & 0 & 0 \\ -1 & 0 & 0 & 0 \end{pmatrix}.$$

### 26. Rotator

Circular retarder with retardance = $2\theta$ .

$$\mathbf{M}=\begin{pmatrix} 1 & 0 & 0 & 0 \\ 0 & \cos 2\theta & \sin 2\theta & 0 \\ 0 & -\sin 2\theta & \cos 2\theta & 0 \\ 0 & 0 & 0 & 1 \end{pmatrix},\mathbf{C}=\begin{pmatrix} 2\cos^2\theta & 0 & 0 & -i\sin 2\theta \\ 0 & 0 & 0 & 0 \\ 0 & 0 & 0 & 0 \\ i\sin 2\theta & 0 & 0 & 2\sin^2\theta \end{pmatrix},$$

$$\mathbf{F}=\sqrt{2}\begin{pmatrix} \cos\theta & 0 & 0 & 0 \\ 0 & 0 & 0 & 0 \\ 0 & 0 & 0 & 0 \\ i\sin\theta & 0 & 0 & 0 \end{pmatrix}.$$

### 27. Elliptic retarder

Phase shift between the two eigenstates = $\Delta$ ,
**T**he ratio between the strengths of components of the electric field **=** tan $\chi$ **,** ellipticity angle = $\chi$ **.** Oriented at an angle $\phi$ .
Terminology as in [1].

$$\mathbf{M}=\frac{1}{4}\begin{pmatrix} 1 & 0 & 0 & 0 \\ 0 & \cos^2\frac{\Delta}{2}-\sin^2\frac{\Delta}{2}(\sin^2 2\chi-\cos^2 2\chi\cos 4\phi) & \sin^2\frac{\Delta}{2}\cos^2 2\chi\sin 4\phi+\sin\Delta\sin 2\chi & \sin^2\frac{\Delta}{2}\sin 4\chi\cos 2\phi-\sin\Delta\cos 2\chi\sin 2\phi \\ 0 & \sin^2\frac{\Delta}{2}\cos^2 2\chi\sin 4\phi-\sin\Delta\sin 2\chi & \cos^2\frac{\Delta}{2}-\sin^2\frac{\Delta}{2}(\sin^2 2\chi+\cos^2 2\chi\cos 4\phi) & \sin^2\frac{\Delta}{2}\sin 4\chi\sin 2\phi+\sin\Delta\cos 2\chi\cos 2\phi \\ 0 & \sin^2\frac{\Delta}{2}\sin 4\chi\cos 2\phi+\sin\Delta\cos 2\chi\sin 2\phi & \sin^2\frac{\Delta}{2}\sin 4\chi\sin 2\phi-\sin\Delta\cos 2\chi\cos 2\phi & \cos^2\frac{\Delta}{2}-\sin^2\frac{\Delta}{2}\cos 4\chi \end{pmatrix}$$

$$\mathbf{C}=\frac{1}{4}\begin{pmatrix} 2\cos^2\frac{\Delta}{2} & -i\sin\Delta\cos 2\chi\cos 2\phi & -i\sin\Delta\cos 2\chi\sin 2\phi & -i\sin\Delta\sin 2\chi \\ i\sin\Delta\cos 2\chi\cos 2\phi & 2\sin^2\frac{\Delta}{2}\cos^2 2\chi\cos^2 2\phi & \sin^2\frac{\Delta}{2}\cos^2 2\chi\sin 4\phi & \sin^2\frac{\Delta}{2}\sin 4\chi\cos 2\phi \\ i\sin\Delta\cos 2\chi\sin 2\phi & \sin^2\frac{\Delta}{2}\cos^2 2\chi\sin 4\phi & 2\sin^2\frac{\Delta}{2}\cos^2 2\chi\sin^2 2\phi & \sin^2\frac{\Delta}{2}\sin 4\chi\sin 2\phi \\ i\sin\Delta\sin 2\chi & \sin^2\frac{\Delta}{2}\sin 4\chi\cos 2\phi & \sin^2\frac{\Delta}{2}\sin 4\chi\sin 2\phi & 2\sin^2\frac{\Delta}{2}\sin^2 2\chi \end{pmatrix},$$

$$\mathbf{F}=\frac{1}{\sqrt{2}}\begin{pmatrix} \cos\frac{\Delta}{2} & 0 & 0 & 0 \\ i\sin\frac{\Delta}{2}\cos 2\chi\cos 2\phi & 0 & 0 & 0 \\ i\sin\frac{\Delta}{2}\cos 2\chi\sin 2\phi & 0 & 0 & 0 \\ i\sin\frac{\Delta}{2}\sin 2\chi & 0 & 0 & 0 \end{pmatrix}.$$

**28. Elliptic retarder at** $0^{\circ}$

Phase shift between the two eigenstates = $\Delta$ ,
**T**he ratio between the strengths of components of the electric field **=** tan $\chi$ **,** ellipticity angle = $\chi$ **.** Terminology as in [1].

$$\mathbf{M}=\begin{pmatrix} 1 & 0 & 0 & 0 \\ 0 & \cos^2 2\chi+\cos\Delta\sin^2 2\chi & \sin\Delta\sin 2\chi & (1-\cos\Delta)\sin 2\chi\cos 2\chi \\ 0 & -\sin\Delta\sin 2\chi & \cos\Delta & \sin\Delta\cos 2\chi \\ 0 & (1-\cos\Delta)\sin 2\chi\cos 2\chi & -\sin\Delta\cos 2\chi & \sin^2 2\chi+\cos\Delta\cos^2 2\chi \end{pmatrix},$$

$$\mathbf{C}=2\begin{pmatrix} \cos^2\frac{\Delta}{2} & -i\sin\frac{\Delta}{2}\cos\frac{\Delta}{2}\cos 2\chi & 0 & -i\sin\frac{\Delta}{2}\cos\frac{\Delta}{2}\sin 2\chi \\ i\sin\frac{\Delta}{2}\cos\frac{\Delta}{2}\cos 2\chi & \sin^2\frac{\Delta}{2}\cos^2 2\chi & 0 & \sin^2\frac{\Delta}{2}\sin 2\chi\cos 2\chi \\ 0 & 0 & 0 & 0 \\ i\sin\frac{\Delta}{2}\cos\frac{\Delta}{2}\sin 2\chi & \sin^2\frac{\Delta}{2}\sin 2\chi\cos 2\chi & 0 & \sin^2\frac{\Delta}{2}\sin^2 2\chi \end{pmatrix},$$

$$\mathbf{F}=\sqrt{2}\begin{pmatrix} \cos\frac{\Delta}{2} & 0 & 0 & 0 \\ i\sin\frac{\Delta}{2}\cos 2\chi & 0 & 0 & 0 \\ 0 & 0 & 0 & 0 \\ i\sin\frac{\Delta}{2}\sin 2\chi & 0 & 0 & 0 \end{pmatrix}.$$

**29. Elliptic retarder at** $45^{\circ}$

$$\mathbf{M}=\frac{1}{4}\begin{pmatrix} 1 & 0 & 0 & 0 \\ 0 & \cos\Delta & \sin\Delta\sin 2\chi & -\sin\Delta\cos 2\chi \\ 0 & -\sin\Delta\sin 2\chi & \cos^2\frac{\Delta}{2}+\sin^2\frac{\Delta}{2}\cos 4\chi & \sin^2\frac{\Delta}{2}\sin 4\chi \\ 0 & \sin\Delta\cos 2\chi & \sin^2\frac{\Delta}{2}\sin 4\chi & \cos^2\frac{\Delta}{2}-\sin^2\frac{\Delta}{2}\cos 4\chi \end{pmatrix}$$

$$\mathbf{C}=\frac{1}{4}\begin{pmatrix} 2\cos^2\frac{\Delta}{2} & 0 & -i\sin\Delta\cos 2\chi & -i\sin\Delta\sin 2\chi \\ 0 & 0 & 0 & 0 \\ i\sin\Delta\cos 2\chi & 0 & 2\sin^2\frac{\Delta}{2}\cos^2 2\chi & \sin^2\frac{\Delta}{2}\sin 4\chi \\ i\sin\Delta\sin 2\chi & 0 & \sin^2\frac{\Delta}{2}\sin 4\chi & 2\sin^2\frac{\Delta}{2} \end{pmatrix},$$

$$\mathbf{F}=\frac{1}{\sqrt{2}}\begin{pmatrix} \cos\frac{\Delta}{2} & 0 & 0 & 0 \\ 0 & 0 & 0 & 0 \\ i\sin\frac{\Delta}{2}\cos 2\chi & 0 & 0 & 0 \\ i\sin\frac{\Delta}{2}\sin 2\chi & 0 & 0 & 0 \end{pmatrix}.$$

**30. Elliptic retarder at** $90^{\circ}$

$$\mathbf{M}=\frac{1}{4}\begin{pmatrix} 1 & 0 & 0 & 0 \\ 0 & \cos^2\frac{\Delta}{2}+\sin^2\frac{\Delta}{2}\cos 4\chi & \sin\Delta\sin 2\chi & -\sin^2\frac{\Delta}{2}\sin 4\chi \\ 0 & -\sin\Delta\sin 2\chi & \cos\Delta & -\sin\Delta\cos 2\chi \\ 0 & -\sin^2\frac{\Delta}{2}\sin 4\chi & \sin\Delta\cos 2\chi & \cos^2\frac{\Delta}{2}-\sin^2\frac{\Delta}{2}\cos 4\chi \end{pmatrix}$$

$$\mathbf{C}=\frac{1}{4}\begin{pmatrix} 2\cos^2\frac{\Delta}{2} & i\sin\Delta\cos 2\chi & 0 & -i\sin\Delta\sin 2\chi \\ -i\sin\Delta\cos 2\chi & 2\sin^2\frac{\Delta}{2}\cos^2 2\chi & 0 & -\sin^2\frac{\Delta}{2}\sin 4\chi \\ 0 & 0 & 0 & 0 \\ i\sin\Delta\sin 2\chi & -\sin^2\frac{\Delta}{2}\sin 4\chi & 0 & 2\sin^2\frac{\Delta}{2}\sin^2 2\chi \end{pmatrix},$$

$$\mathbf{F}=\frac{1}{\sqrt{2}}\begin{pmatrix} \cos\frac{\Delta}{2} & 0 & 0 & 0 \\ -i\sin\frac{\Delta}{2}\cos 2\chi & 0 & 0 & 0 \\ 0 & 0 & 0 & 0 \\ i\sin\frac{\Delta}{2}\sin 2\chi & 0 & 0 & 0 \end{pmatrix}.$$

**31. Elliptic retarder at** $135^{\circ}$

$$\mathbf{M}=\frac{1}{4}\begin{pmatrix} 1 & 0 & 0 & 0 \\ 0 & \cos\Delta & \sin\Delta\sin 2\chi & \sin\Delta\cos 2\chi \\ 0 & -\sin\Delta\sin 2\chi & \cos^2\frac{\Delta}{2}+\sin^2\frac{\Delta}{2}\cos 4\chi & -\sin^2\frac{\Delta}{2}\sin 4\chi \\ 0 & -\sin\Delta\cos 2\chi & -\sin^2\frac{\Delta}{2}\sin 4\chi & \cos^2\frac{\Delta}{2}-\sin^2\frac{\Delta}{2}\cos 4\chi \end{pmatrix}$$

$$\mathbf{C}=\frac{1}{4}\begin{pmatrix} 2\cos^2\frac{\Delta}{2} & 0 & i\sin\Delta\cos 2\chi & -i\sin\Delta\sin 2\chi \\ 0 & 0 & 0 & 0 \\ -i\sin\Delta\cos 2\chi & 0 & 2\sin^2\frac{\Delta}{2}\cos^2 2\chi & -\sin^2\frac{\Delta}{2}\sin 4\chi \\ i\sin\Delta\sin 2\chi & 0 & -\sin^2\frac{\Delta}{2}\sin 4\chi & 2\sin^2\frac{\Delta}{2}\sin^2 2\chi \end{pmatrix},$$

$$\mathbf{F}=\frac{1}{\sqrt{2}}\begin{pmatrix} \cos\frac{\Delta}{2} & 0 & 0 & 0 \\ 0 & 0 & 0 & 0 \\ -i\sin\frac{\Delta}{2}\cos 2\chi & 0 & 0 & 0 \\ i\sin\frac{\Delta}{2}\sin 2\chi & 0 & 0 & 0 \end{pmatrix}.$$

**32. Linear retarder**

Phase shift $\Delta$ ,
Oriented at an angle $\phi$ .Terminology as in [1].

$$\mathbf{M}=\begin{pmatrix} 1 & 0 & 0 & 0 \\ 0 & \cos^2 2\phi+\cos\Delta\sin^2 2\phi & (1-\cos\Delta)\sin 2\phi\cos 2\phi & -\sin\Delta\sin 2\phi \\ 0 & (1-\cos\Delta)\sin 2\phi\cos 2\phi & \sin^2 2\phi+\cos\Delta\cos^2 2\phi & \sin\Delta\cos 2\phi \\ 0 & \sin\Delta\sin 2\phi & -\sin\Delta\cos 2\phi & \cos\Delta \end{pmatrix},$$

$$\mathbf{C}=2\begin{pmatrix} \cos^2\frac{\Delta}{2} & -i\sin\frac{\Delta}{2}\cos\frac{\Delta}{2}\cos 2\phi & -i\sin\frac{\Delta}{2}\cos\frac{\Delta}{2}\sin 2\phi & 0 \\ i\sin\frac{\Delta}{2}\cos\frac{\Delta}{2}\cos 2\phi & \sin^2\frac{\Delta}{2}\cos^2 2\phi & \sin^2\frac{\Delta}{2}\sin 2\phi\cos 2\phi & 0 \\ i\sin\frac{\Delta}{2}\cos\frac{\Delta}{2}\sin 2\phi & \sin^2\frac{\Delta}{2}\sin 2\phi\cos 2\phi & \sin^2\frac{\Delta}{2}\sin^2 2\phi & 0 \\ 0 & 0 & 0 & 0 \end{pmatrix},$$

$$\mathbf{F}=\sqrt{2}\begin{pmatrix} \cos\frac{\Delta}{2} & 0 & 0 & 0 \\ i\sin\frac{\Delta}{2}\cos 2\phi & 0 & 0 & 0 \\ i\sin\frac{\Delta}{2}\sin 2\phi & 0 & 0 & 0 \\ 0 & 0 & 0 & 0 \end{pmatrix}.$$

**33. Linear retarder at** $0^\circ$

Phase shift $\Delta$

$$\mathbf{M}=\begin{pmatrix} 1 & 0 & 0 & 0 \\ 0 & 1 & 0 & 0 \\ 0 & 0 & \cos\Delta & \sin\Delta \\ 0 & 0 & -\sin\Delta & \cos\Delta \end{pmatrix},\mathbf{C}=2\begin{pmatrix} \cos^2\frac{\Delta}{2} & -i\sin\frac{\Delta}{2}\cos\frac{\Delta}{2} & 0 & 0 \\ i\sin\frac{\Delta}{2}\cos\frac{\Delta}{2} & \sin^2\frac{\Delta}{2} & 0 & 0 \\ 0 & 0 & 0 & 0 \\ 0 & 0 & 0 & 0 \end{pmatrix},$$

$$\mathbf{F}=\sqrt{2}\begin{pmatrix} \cos\frac{\Delta}{2} & 0 & 0 & 0 \\ i\sin\frac{\Delta}{2} & 0 & 0 & 0 \\ 0 & 0 & 0 & 0 \\ 0 & 0 & 0 & 0 \end{pmatrix}.$$

**34. Linear retarder at** $45^\circ$

$$\mathbf{M}=\begin{pmatrix} 1 & 0 & 0 & 0 \\ 0 & \cos\Delta & 0 & -\sin\Delta \\ 0 & 0 & 1 & 0 \\ 0 & \sin\Delta & 0 & \cos\Delta \end{pmatrix},$$

$$\mathbf{C}=2\begin{pmatrix} \cos^2\frac{\Delta}{2} & 0 & -i\sin\frac{\Delta}{2}\cos\frac{\Delta}{2} & 0 \\ 0 & 0 & 0 & 0 \\ i\sin\frac{\Delta}{2}\cos\frac{\Delta}{2} & 0 & \sin^2\frac{\Delta}{2} & 0 \\ 0 & 0 & 0 & 0 \end{pmatrix},$$

$$\mathbf{F}=\sqrt{2}\begin{pmatrix} \cos\frac{\Delta}{2} & 0 & 0 & 0 \\ 0 & 0 & 0 & 0 \\ i\sin\frac{\Delta}{2} & 0 & 0 & 0 \\ 0 & 0 & 0 & 0 \end{pmatrix}.$$

**35. Linear retarder at** $90^\circ$

$$\mathbf{M}=\begin{pmatrix} 1 & 0 & 0 & 0 \\ 0 & 1 & 0 & 0 \\ 0 & 0 & \cos\Delta & -\sin\Delta \\ 0 & 0 & \sin\Delta & \cos\Delta \end{pmatrix},$$

$$\mathbf{C}=2\begin{pmatrix} \cos^2\frac{\Delta}{2} & i\sin\frac{\Delta}{2}\cos\frac{\Delta}{2} & 0 & 0 \\ -i\sin\frac{\Delta}{2}\cos\frac{\Delta}{2} & \sin^2\frac{\Delta}{2} & 0 & 0 \\ 0 & 0 & 0 & 0 \\ 0 & 0 & 0 & 0 \end{pmatrix},$$

$$\mathbf{F}=\sqrt{2}\begin{pmatrix} \cos\frac{\Delta}{2} & 0 & 0 & 0 \\ -i\sin\frac{\Delta}{2} & 0 & 0 & 0 \\ 0 & 0 & 0 & 0 \\ 0 & 0 & 0 & 0 \end{pmatrix}.$$

**36. Linear retarder at** $135^\circ$

$$\mathbf{M}=\begin{pmatrix} 1 & 0 & 0 & 0 \\ 0 & \cos\Delta & 0 & \sin\Delta \\ 0 & 0 & 1 & 0 \\ 0 & -\sin\Delta & 0 & \cos\Delta \end{pmatrix},$$

$$\mathbf{C} = 2\begin{pmatrix} \cos^2\frac{\Delta}{2} & 0 & i\sin\frac{\Delta}{2}\cos\frac{\Delta}{2} & 0 \\ 0 & 0 & 0 & 0 \\ -i\sin\frac{\Delta}{2}\cos\frac{\Delta}{2} & 0 & \sin^2\frac{\Delta}{2} & 0 \\ 0 & 0 & 0 & 0 \end{pmatrix},$$

$$\mathbf{F} = \sqrt{2}\begin{pmatrix} \cos\frac{\Delta}{2} & 0 & 0 & 0 \\ 0 & 0 & 0 & 0 \\ -i\sin\frac{\Delta}{2} & 0 & 0 & 0 \\ 0 & 0 & 0 & 0 \end{pmatrix}.$$

**37. Quarter-wave retarder at** $0^\circ$

$$\mathbf{M} = \begin{pmatrix} 1 & 0 & 0 & 0 \\ 0 & 1 & 0 & 0 \\ 0 & 0 & 0 & 1 \\ 0 & 0 & -1 & 0 \end{pmatrix}, \mathbf{C} = \begin{pmatrix} 1 & -i & 0 & 0 \\ i & 1 & 0 & 0 \\ 0 & 0 & 0 & 0 \\ 0 & 0 & 0 & 0 \end{pmatrix}, \mathbf{F} = \begin{pmatrix} 1 & 0 & 0 & 0 \\ i & 0 & 0 & 0 \\ 0 & 0 & 0 & 0 \\ 0 & 0 & 0 & 0 \end{pmatrix}.$$

**38. Quarter-wave retarder at** $45^\circ$

$$\mathbf{M} = \begin{pmatrix} 1 & 0 & 0 & 0 \\ 0 & 0 & 0 & -1 \\ 0 & 0 & 1 & 0 \\ 0 & 1 & 0 & 0 \end{pmatrix}, \mathbf{C} = \begin{pmatrix} 1 & 0 & -i & 0 \\ 0 & 0 & 0 & 0 \\ i & 0 & 1 & 0 \\ 0 & 0 & 0 & 0 \end{pmatrix}, \mathbf{F} = \begin{pmatrix} 1 & 0 & 0 & 0 \\ 0 & 0 & 0 & 0 \\ i & 0 & 0 & 0 \\ 0 & 0 & 0 & 0 \end{pmatrix}.$$

**39. Quarter-wave retarder at** $90^\circ$

$$\mathbf{M} = \begin{pmatrix} 1 & 0 & 0 & 0 \\ 0 & 1 & 0 & 0 \\ 0 & 0 & 0 & -1 \\ 0 & 0 & 1 & 0 \end{pmatrix}, \mathbf{C} = \begin{pmatrix} 1 & i & 0 & 0 \\ -i & 1 & 0 & 0 \\ 0 & 0 & 0 & 0 \\ 0 & 0 & 0 & 0 \end{pmatrix}, \mathbf{F} = \begin{pmatrix} 1 & 0 & 0 & 0 \\ -i & 0 & 0 & 0 \\ 0 & 0 & 0 & 0 \\ 0 & 0 & 0 & 0 \end{pmatrix}.$$

**40. Quarter-wave retarder at** $135^\circ$

$$\mathbf{M} = \begin{pmatrix} 1 & 0 & 0 & 0 \\ 0 & 0 & 0 & 1 \\ 0 & 0 & 1 & 0 \\ 0 & -1 & 0 & 0 \end{pmatrix}, \mathbf{C} = \begin{pmatrix} 1 & 0 & i & 0 \\ 0 & 0 & 0 & 0 \\ -i & 0 & 1 & 0 \\ 0 & 0 & 0 & 0 \end{pmatrix}, \mathbf{F} = \begin{pmatrix} 1 & 0 & 0 & 0 \\ 0 & 0 & 0 & 0 \\ -i & 0 & 0 & 0 \\ 0 & 0 & 0 & 0 \end{pmatrix}.$$

**41. Circular retarder**

Phase shift $\Delta$.

$$\mathbf{M}=\begin{pmatrix}1&0&0&0\\0&\cos\Delta&\sin\Delta&0\\0&-\sin\Delta&\cos\Delta&0\\0&0&0&1\end{pmatrix},\mathbf{C}=2\begin{pmatrix}\cos^2\frac{\Delta}{2}&0&0&-i\sin\frac{\Delta}{2}\cos\frac{\Delta}{2}\\0&0&0&0\\0&0&0&0\\i\sin\frac{\Delta}{2}\cos\frac{\Delta}{2}&0&0&\sin^2\frac{\Delta}{2}\end{pmatrix},$$

$$\mathbf{F}=\sqrt{2}\begin{pmatrix}\cos\frac{\Delta}{2}&0&0&0\\0&0&0&0\\0&0&0&0\\i\sin\frac{\Delta}{2}&0&0&0\end{pmatrix}.$$

**42. Quarter-wave retarder circular left**

$$\mathbf{M}=\begin{pmatrix}1&0&0&0\\0&0&1&0\\0&-1&0&0\\0&0&0&1\end{pmatrix},\mathbf{C}=\begin{pmatrix}1&0&0&-i\\0&0&0&0\\0&0&0&0\\i&0&0&1\end{pmatrix},\mathbf{F}=\begin{pmatrix}1&0&0&0\\0&0&0&0\\0&0&0&0\\i&0&0&0\end{pmatrix}.$$

**43. Quarter-wave retarder circular right**

$$\mathbf{M}=\begin{pmatrix}1&0&0&i\\0&0&0&0\\0&0&0&0\\-i&0&0&1\end{pmatrix},\mathbf{C}=\begin{pmatrix}1&0&0&i\\0&0&0&0\\0&0&0&0\\-i&0&0&1\end{pmatrix},\mathbf{F}=\begin{pmatrix}1&0&0&0\\0&0&0&0\\0&0&0&0\\-i&0&0&0\end{pmatrix}.$$

**44. Half-wave retarder (pseudorotator)**

Oriented at an angle $\phi$.

$$\mathbf{M}=\begin{pmatrix}1&0&0&0\\0&\cos4\phi&\sin4\phi&0\\0&\sin4\phi&-\cos4\phi&0\\0&0&0&-1\end{pmatrix},\mathbf{C}=\begin{pmatrix}0&0&0&0\\0&1+\cos4\phi&\sin4\phi&0\\0&\sin4\phi&1-\cos4\phi&0\\0&0&0&0\end{pmatrix},$$

$$\mathbf{F}=\sqrt{2}\begin{pmatrix}0&0&0&0\\\cos2\phi&0&0&0\\\sin2\phi&0&0&0\\0&0&0&0\end{pmatrix}.$$

**45. Half-wave retarder $0^\circ$**

$$\mathbf{M}=\begin{pmatrix}1&0&0&0\\0&1&0&0\\0&0&-1&0\\0&0&0&-1\end{pmatrix},\mathbf{C}=2\begin{pmatrix}0&0&0&0\\0&1&0&0\\0&0&0&0\\0&0&0&0\end{pmatrix},\mathbf{F}=\sqrt{2}\begin{pmatrix}0&0&0&0\\1&0&0&0\\0&0&0&0\\0&0&0&0\end{pmatrix}.$$

**46. Half-wave retarder** $45^\circ$

$$\mathbf{M}=\begin{pmatrix}1&0&0&0\\0&-1&0&0\\0&0&1&0\\0&0&0&-1\end{pmatrix},\mathbf{C}=2\begin{pmatrix}0&0&0&0\\0&0&0&0\\0&0&1&0\\0&0&0&0\end{pmatrix},\mathbf{F}=\sqrt{2}\begin{pmatrix}0&0&0&0\\0&0&0&0\\1&0&0&0\\0&0&0&0\end{pmatrix}.$$

**47. Half-wave retarder** $90^\circ$ **(** $\mathbf{M},\mathbf{C}$ **are same as for** $0^\circ$ **)**

$$\mathbf{M}=\begin{pmatrix}1&0&0&0\\0&1&0&0\\0&0&-1&0\\0&0&0&-1\end{pmatrix},\mathbf{C}=2\begin{pmatrix}0&0&0&0\\0&1&0&0\\0&0&0&0\\0&0&0&0\end{pmatrix},\mathbf{F}=\sqrt{2}\begin{pmatrix}0&0&0&0\\-1&0&0&0\\0&0&0&0\\0&0&0&0\end{pmatrix}.$$

**48. Half-wave retarder** $135^\circ$

$$\mathbf{M}=\begin{pmatrix}1&0&0&0\\0&-1&0&0\\0&0&1&0\\0&0&0&-1\end{pmatrix},\mathbf{C}=2\begin{pmatrix}0&0&0&0\\0&0&0&0\\0&0&1&0\\0&0&0&0\end{pmatrix},\mathbf{F}=\sqrt{2}\begin{pmatrix}0&0&0&0\\0&0&0&0\\-1&0&0&0\\0&0&0&0\end{pmatrix}.$$

**49. Normal reflection from a dielectric**

Refractive index $= n$

$$\mathbf{M}=\left(\frac{n-1}{n+1}\right)^2\begin{pmatrix}1&0&0&0\\0&1&0&0\\0&0&-1&0\\0&0&0&-1\end{pmatrix},\mathbf{C}=2\left(\frac{n-1}{n+1}\right)^2\begin{pmatrix}0&0&0&0\\0&1&0&0\\0&0&0&0\\0&0&0&0\end{pmatrix},$$

$$\mathbf{F}=\sqrt{2}\left(\frac{n-1}{n+1}\right)\begin{pmatrix}0&0&0&0\\-1&0&0&0\\0&0&0&0\\0&0&0&0\end{pmatrix}.$$

**50. Half-wave retarder** $45^\circ$

$$\mathbf{M}=\begin{pmatrix}1&0&0&0\\0&-1&0&0\\0&0&1&0\\0&0&0&-1\end{pmatrix},\mathbf{C}=2\begin{pmatrix}0&0&0&0\\0&0&0&0\\0&0&1&0\\0&0&0&0\end{pmatrix},\mathbf{F}=\sqrt{2}\begin{pmatrix}0&0&0&0\\0&0&0&0\\1&0&0&0\\0&0&0&0\end{pmatrix}.$$

**51. Half-wave retarder** $135^\circ$

$$\mathbf{M}=\begin{pmatrix}1&0&0&0\\0&-1&0&0\\0&0&1&0\\0&0&0&-1\end{pmatrix},\mathbf{C}=2\begin{pmatrix}0&0&0&0\\0&0&0&0\\0&0&1&0\\0&0&0&0\end{pmatrix},\mathbf{F}=\sqrt{2}\begin{pmatrix}0&0&0&0\\0&0&0&0\\-1&0&0&0\\0&0&0&0\end{pmatrix}.$$

**52. Half-wave retarder left**

$$\mathbf{M}=\begin{pmatrix} 1 & 0 & 0 & 0 \\ 0 & -1 & 0 & 0 \\ 0 & 0 & -1 & 0 \\ 0 & 0 & 0 & 1 \end{pmatrix},\mathbf{C}=2\begin{pmatrix} 0 & 0 & 0 & 0 \\ 0 & 0 & 0 & 0 \\ 0 & 0 & 0 & 0 \\ 0 & 0 & 0 & 1 \end{pmatrix},\mathbf{F}=\sqrt{2}\begin{pmatrix} 0 & 0 & 0 & 0 \\ 0 & 0 & 0 & 0 \\ 0 & 0 & 0 & 0 \\ 1 & 0 & 0 & 0 \end{pmatrix}.$$

**53. Half-wave retarder right**

$$\mathbf{M}=\begin{pmatrix} 1 & 0 & 0 & 0 \\ 0 & -1 & 0 & 0 \\ 0 & 0 & -1 & 0 \\ 0 & 0 & 0 & 1 \end{pmatrix},\mathbf{C}=2\begin{pmatrix} 0 & 0 & 0 & 0 \\ 0 & 0 & 0 & 0 \\ 0 & 0 & 0 & 0 \\ 0 & 0 & 0 & 1 \end{pmatrix},\mathbf{F}=\sqrt{2}\begin{pmatrix} 0 & 0 & 0 & 0 \\ 0 & 0 & 0 & 0 \\ 0 & 0 & 0 & 0 \\ -1 & 0 & 0 & 0 \end{pmatrix}.$$

**54. Aligned linear diattenuation and retardance**

$$\mathbf{M}=\begin{pmatrix} A & B & 0 & 0 \\ B & A & 0 & 0 \\ 0 & 0 & C & D \\ 0 & 0 & -D & C \end{pmatrix},\mathbf{C}=\begin{pmatrix} A+C & B-iD & 0 & 0 \\ B+iD & A-C & 0 & 0 \\ 0 & 0 & 0 & 0 \\ 0 & 0 & 0 & 0 \end{pmatrix},$$

$$\mathbf{F}=\frac{1}{\sqrt{2R}}\begin{pmatrix} (R+C)\sqrt{(R-C)(A+R)} & -(R-C)\sqrt{(R+C)(A-R)} & 0 & 0 \\ \sqrt{\dfrac{(A+R)}{(R+C)}}(B+iD) & \sqrt{\dfrac{(A-R)}{(R-C)}}(B+iD) & 0 & 0 \\ 0 & 0 & 0 & 0 \\ 0 & 0 & 0 & 0 \end{pmatrix},$$

$$A\geq R, R=\sqrt{B^2+C^2+D^2}.$$

**55. Medium of scattering particles obeying reciprocity and with a plane of symmetry [2]**

$$\mathbf{M}=\begin{pmatrix} A_0+B_0 & C & 0 & 0 \\ C & A+B & 0 & 0 \\ 0 & 0 & A-B & D \\ 0 & 0 & -D & A_0-B_0 \end{pmatrix},\mathbf{C}=\begin{pmatrix} A_0+A & C-iD & 0 & 0 \\ C+iD & B_0+B & 0 & 0 \\ 0 & 0 & B_0-B & 0 \\ 0 & 0 & 0 & A_0-A \end{pmatrix},$$

$$A_0\geq A\geq 0, B_0\geq B\geq 0.$$

**56. Rotationally symetric medium of asymmetric scatterers in the direct forward scattering direction [2]**

$$\mathbf{M}=\begin{pmatrix} A_0+B_0 & 0 & 0 & F+I \\ 0 & A & J & 0 \\ 0 & -J & A & 0 \\ I-F & 0 & 0 & A_0-B_0 \end{pmatrix},\mathbf{C}=\begin{pmatrix} A_0+A & 0 & 0 & I-iJ \\ 0 & B_0 & iF & 0 \\ 0 & -iF & B_0 & 0 \\ I+iJ & 0 & 0 & A_0-A \end{pmatrix},$$

$$\mathbf{F}=\frac{1}{\sqrt{2}}\begin{pmatrix} \dfrac{(R+A)\sqrt{(R-A)(A_0+R)}}{R} & -\dfrac{(R-A)\sqrt{(R+A)(A_0-R)}}{R} & 0 & 0 \\ 0 & 0 & \sqrt{B_0+F} & \sqrt{B_0-F} \\ 0 & 0 & -i\sqrt{B_0+F} & i\sqrt{B_0-F} \\ \sqrt{\dfrac{(A_0+R)}{R(R+A)}}(I+iJ) & \sqrt{\dfrac{(A_0-R)}{R(R-A)}}(I+iJ) & 0 & 0 \end{pmatrix},$$

$$A_0\geq R\geq 0, B_0\geq F, R=\sqrt{A^2+I^2+J^2}.$$

**57. Medium of asymmetric scatterers in the exact backscattering direction [2]**

$$\mathbf{M}=\begin{pmatrix} A_0+B & C & L & I \\ C & A+B & J & K \\ -L & -J & A-B & D \\ I & K & -D & A_0-B \end{pmatrix},\mathbf{C}=\begin{pmatrix} A_0+A & C-iD & 0 & I-iJ \\ C+iD & 2B & 0 & K-iL \\ 0 & 0 & 0 & 0 \\ I+iJ & K+iL & 0 & A_0-A \end{pmatrix},$$

$$A_0\ge A\ge 0, B\ge 0.$$

**58. Medium of asymmetric scatterers in the non-exact backscattering direction, with rotational symmetry of the medium [2]**

$$\mathbf{M}=\begin{pmatrix} A_1+B & 0 & 0 & I \\ 0 & B & 0 & 0 \\ 0 & 0 & -B & 0 \\ I & 0 & 0 & A_2-B \end{pmatrix},\mathbf{C}=\frac{1}{2}\begin{pmatrix} A_1+A_2 & 0 & 0 & 2I \\ 0 & 4B+A_1-A_2 & 0 & 0 \\ 0 & 0 & A_1-A_2 & 0 \\ 2I & 0 & 0 & A_1+A_2 \end{pmatrix},$$

$$\mathbf{F}=\frac{1}{2}\begin{pmatrix} \sqrt{A_1+A_2+2I} & \sqrt{A_1+A_2-2I} & 0 & 0 \\ 0 & 0 & -\sqrt{2(4B+A_1-A_2)} & 0 \\ 0 & 0 & 0 & \sqrt{2(A_1-A_2)} \\ \sqrt{A_1+A_2+2I} & -\sqrt{A_1+A_2-2I} & 0 & 0 \end{pmatrix},$$

$$A_1\ge A_2\ge 0, B\ge 4(A_2-A_1).$$

**59. Medium of asymmetric scatterers in the exact backscattering direction, with rotational symmetry of the medium [2]**

$$\mathbf{M}=\begin{pmatrix} A_0+B & 0 & 0 & I \\ 0 & B & 0 & 0 \\ 0 & 0 & -B & 0 \\ I & 0 & 0 & A_0-B \end{pmatrix},\mathbf{C}=\begin{pmatrix} A_0 & 0 & 0 & I \\ 0 & 2B & 0 & 0 \\ 0 & 0 & 0 & 0 \\ I & 0 & 0 & A_0 \end{pmatrix},$$

$$\mathbf{F}=\frac{1}{\sqrt{2}}\begin{pmatrix} \sqrt{A_0+I} & \sqrt{A_0-I} & 0 & 0 \\ 0 & 0 & -2\sqrt{B} & 0 \\ 0 & 0 & 0 & 0 \\ \sqrt{A_0+I} & -\sqrt{A_0-I} & 0 & 0 \end{pmatrix}, A_0\ge 0, B\ge 0.$$

**60. G-antisymmetric Mueller matrix [3]**

$$\mathbf{M}=\begin{pmatrix} a & p_1 & p_2 & p_3 \\ p_1 & a & p_6 & -p_5 \\ p_2 & -p_6 & a & p_4 \\ p_3 & p_5 & -p_4 & a \end{pmatrix},\mathbf{C}=\begin{pmatrix} 2a & p_1-ip_4 & p_2-ip_5 & p_3-ip_6 \\ p_1+ip_4 & 0 & 0 & 0 \\ p_2+ip_5 & 0 & 0 & 0 \\ p_3+ip_6 & 0 & 0 & 0 \end{pmatrix},$$

$$\mathbf{F}=\begin{pmatrix} (a+r)^{3/2} & (a-r)^{3/2} & 0 & 0 \\ \sqrt{a+r}(p_1+ip_4) & \sqrt{a-r}(p_1+ip_4) & 0 & 0 \\ \sqrt{a+r}(p_2+ip_5) & \sqrt{a-r}(p_2+ip_5) & 0 & 0 \\ \sqrt{a+r}(p_3+ip_6) & \sqrt{a-r}(p_3+ip_6) & 0 & 0 \end{pmatrix}.$$

$$r=\sqrt{a^2+p_1^2+p_2^2+p_3^2+p_4^2+p_5^2+p_6^2}, a\ge r.$$

**61. G-symmetric Mueller matrix [3]**

$$\mathbf{M}=\begin{pmatrix} 0 & p_1 & p_2 & p_3 \\ -p_1 & 0 & p_6 & p_5 \\ -p_2 & p_6 & 0 & p_4 \\ -p_3 & p_5 & p_4 & 0 \end{pmatrix},\mathbf{C}=\begin{pmatrix} 0 & 0 & 0 & 0 \\ 0 & 0 & p_6+ip_3 & p_5-ip_2 \\ 0 & p_6-ip_3 & 0 & p_4+ip_1 \\ 0 & p_5+ip_2 & p_4-ip_1 & 0 \end{pmatrix}.$$

**62. Diagonal Mueller matrix (canonical Mueller matrix,Type1)**

$$\mathbf{M}=\begin{pmatrix} d_0+d_1+d_2+d_3 & 0 & 0 & 0 \\ 0 & d_0+d_1-d_2-d_3 & 0 & 0 \\ 0 & 0 & d_0-d_1+d_2-d_3 & 0 \\ 0 & 0 & 0 & d_0-d_1-d_2+d_3 \end{pmatrix},$$

$$\mathbf{C}=2\begin{pmatrix} d_0 & 0 & 0 & 0 \\ 0 & d_1 & 0 & 0 \\ 0 & 0 & d_2 & 0 \\ 0 & 0 & 0 & d_3 \end{pmatrix},\mathbf{F}=\sqrt{2}\begin{pmatrix} \sqrt{d_0} & 0 & 0 & 0 \\ 0 & \sqrt{d_1} & 0 & 0 \\ 0 & 0 & \sqrt{d_2} & 0 \\ 0 & 0 & 0 & \sqrt{d_3} \end{pmatrix}.$$

**63. Canonical Mueller matrix, Type 2 (Ossikovski)[4]**

$$\mathbf{M}=\begin{pmatrix} 2d_0 & -d_0 & 0 & 0 \\ d_0 & 0 & 0 & 0 \\ 0 & 0 & d_2 & 0 \\ 0 & 0 & 0 & d_2 \end{pmatrix},\mathbf{C}=\begin{pmatrix} d_0+d_2 & 0 & 0 & 0 \\ 0 & d_0-d_2 & 0 & 0 \\ 0 & 0 & d_0 & -id_0 \\ 0 & 0 & id_0 & d_0 \end{pmatrix},$$

$$\mathbf{F}=\begin{pmatrix} \sqrt{d_0+d_2} & 0 & 0 & 0 \\ 0 & \sqrt{d_0-d_2} & 0 & 0 \\ 0 & 0 & -i\sqrt{d_0} & 0 \\ 0 & 0 & \sqrt{d_0} & 0 \end{pmatrix}.$$

**64. Canonical Mueller matrix, Type 2 (Bolshakov) [5, 6]**

$$\mathbf{M}=\begin{pmatrix} 2d_0 & d_0-d_1 & 0 & 0 \\ d_1-d_0 & 2d_1 & 0 & 0 \\ 0 & 0 & d_2 & 0 \\ 0 & 0 & 0 & d_2 \end{pmatrix},\mathbf{C}=\begin{pmatrix} d_0+d_1+d_2 & 0 & 0 & 0 \\ 0 & d_0+d_1-d_2 & 0 & 0 \\ 0 & 0 & d_0-d_1 & i(d_0-d_1) \\ 0 & 0 & -i(d_0-d_1) & d_0-d_1 \end{pmatrix},$$

$$\mathbf{F}=\begin{pmatrix} \sqrt{d_0+d_1+d_2} & 0 & 0 & 0 \\ 0 & \sqrt{d_0+d_1-d_2} & 0 & 0 \\ 0 & 0 & i\sqrt{d_0-d_1} & 0 \\ 0 & 0 & \sqrt{d_0-d_1} & 0 \end{pmatrix},d_0\geq d_1,\left|d_0+d_1\right|\geq\left|d_2\right|.$$

**65. Parameterized deterministic mueller matrix [7]**

$$\mathbf{M} = M_{00}\begin{pmatrix} 1 & l\sin\phi\cos\theta_1 + 2mn\sin^2\frac{\phi}{2}\sin\Delta\theta_1 & m\sin\phi\cos\theta_2 + 2nl\sin^2\frac{\phi}{2}\sin\Delta\theta_2 & n\sin\phi\cos\theta_3 + 2lm\sin^2\frac{\phi}{2}\sin\Delta\theta_3 \\ l\sin\phi\cos\theta_1 - 2mn\sin^2\frac{\phi}{2}\sin\Delta\theta_1 & \cos\phi + 2l^2\sin^2\frac{\phi}{2} & n\sin\phi\cos\theta_3 + 2lm\sin^2\frac{\phi}{2}\cos\Delta\theta_3 & -m\sin\phi\cos\theta_2 + 2nl\sin^2\frac{\phi}{2}\cos\Delta\theta_2 \\ m\sin\phi\cos\theta_2 - 2nl\sin^2\frac{\phi}{2}\sin\Delta\theta_2 & -n\sin\phi\sin\theta_3 + 2lm\sin^2\frac{\phi}{2}\cos\Delta\theta_3 & \cos\phi + 2m^2\sin^2\frac{\phi}{2} & l\sin\phi\cos\theta_1 + 2mn\sin^2\frac{\phi}{2}\cos\Delta\theta_1 \\ n\sin\phi\cos\theta_3 - 2lm\sin^2\frac{\phi}{2}\sin\Delta\theta_3 & m\sin\phi\sin\theta_2 + 2nl\sin^2\frac{\phi}{2}\cos\Delta\theta_2 & -l\sin\phi\sin\theta_1 + 2mn\sin^2\frac{\phi}{2}\cos\Delta\theta_1 & \cos\phi + 2n^2\sin^2\frac{\phi}{2} \end{pmatrix}$$

$l^2 + m^2 + n^2 = 1$, $\Delta\theta_i = (\theta_{i+1} - \theta_{i-1})$ in cyclic order, so that $\Delta\theta_1 + \Delta\theta_2 + \Delta\theta_3 = 0$.

$$\mathbf{C} = 2M_{00}\begin{pmatrix} \cos^2\frac{\phi}{2} & \frac{1}{2}l\sin\phi e^{-i\theta_1} & \frac{1}{2}m\sin\phi e^{-i\theta_2} & \frac{1}{2}n\sin\phi e^{-i\theta_3} \\ \frac{1}{2}l\sin\phi e^{i\theta_1} & l^2\sin^2\frac{\phi}{2} & lm\sin^2\frac{\phi}{2}e^{-i\Delta\theta_3} & nl\sin^2\frac{\phi}{2}e^{i\Delta\theta_2} \\ \frac{1}{2}m\sin\phi e^{i\theta_2} & lm\sin^2\frac{\phi}{2}e^{i\Delta\theta_3} & m^2\sin^2\frac{\phi}{2} & mn\sin^2\frac{\phi}{2}e^{-i\Delta\theta_1} \\ \frac{1}{2}n\sin\phi e^{i\theta_3} & nl\sin^2\frac{\phi}{2}e^{-i\Delta\theta_2} & mn\sin^2\frac{\phi}{2}e^{i\Delta\theta_1} & n^2\sin^2\frac{\phi}{2} \end{pmatrix},$$

$$\mathbf{F} = \sqrt{2M_{00}}\begin{pmatrix} \cos\frac{\phi}{2} & 0 & 0 & 0 \\ l\sin\frac{\phi}{2}e^{i\theta_1} & 0 & 0 & 0 \\ m\sin\frac{\phi}{2}e^{i\theta_2} & 0 & 0 & 0 \\ n\sin\frac{\phi}{2}e^{i\theta_3} & 0 & 0 & 0 \end{pmatrix}.$$

**66. Uniform deterministic medium [8, 9]**

$L = LB - iLD$, $L' = LB' - iLD'$ and $C = CB - iCD$, are the generalized retardances of the medium, where the linear diattenuation in the *x-y* and $45^\circ$ directions are $LD$ and $LD'$, respectively, the circular diattenuation is $CD$, the corresponding retardances are $LB, LB', CB$, all real and $[\mathrm{m}^{-1}]$, the amplitude absorption parameter is $\alpha/2[\mathrm{m}^{-1}]$

$\mathbf{T} = (L, L', -C)^T$, $\mathbf{T}\cdot\mathbf{T}^* = |L|^2 + |L'|^2 + |C|^2$, $T = \left(L^2 + L'^2 + C^2\right)^{1/2} = TB - iTD$.

$\tau = -\arg Tz, \tan\tau = \dfrac{TD}{TB}, \beta = -\arg\tan\left(\frac{1}{2}Tz\right)$, $\dfrac{|T|^2}{\left(\mathbf{T}\cdot\mathbf{T}^*\right)} = \dfrac{\left|\tan(\frac{1}{2}Tz)\right|^2}{\tan^2(\phi/2)} = |p|^2$. $|p|$ is the homogeneity. $\kappa = \arg\cos\left(\frac{1}{2}Tz\right)$ is the geometric phase.

$$\kappa - \beta + \tau = \arg\left(\frac{\sin\left(\frac{1}{2}Tz\right)}{\frac{1}{2}Tz}\right).$$

$$M_{00} = e^{-\alpha z}\left\{\left|\cos\left(\tfrac{1}{2}Tz\right)\right|^2 + \frac{\mathbf{T}\cdot\mathbf{T}^*}{|T|^2}\left|\sin\left(\tfrac{1}{2}Tz\right)\right|^2\right\},$$

$$M_{11} = e^{-\alpha z}\left\{\left|\cos\left(\tfrac{1}{2}Tz\right)\right|^2 - \frac{\left|\sin\left(\frac{1}{2}Tz\right)\right|^2}{|T|^2}\left(\mathbf{T}\cdot\mathbf{T}^* - 2|L|^2\right)\right\},$$

$$M_{22} = e^{-\alpha z}\left\{\left|\cos\left(\tfrac{1}{2}Tz\right)\right|^2 - \frac{\left|\sin\left(\frac{1}{2}Tz\right)\right|^2}{|T|^2}\left(\mathbf{T}\cdot\mathbf{T}^* - 2|L'|^2\right)\right\},$$

$$M_{33} = e^{-\alpha z}\left\{\left|\cos\left(\tfrac{1}{2}Tz\right)\right|^2 - \frac{\left|\sin\left(\frac{1}{2}Tz\right)\right|^2}{|T|^2}\left(\mathbf{T}\cdot\mathbf{T}^* - 2|C|^2\right)\right\}.$$

$$\left.\begin{matrix} M_{01} \\ M_{10} \end{matrix}\right\} = -e^{-\alpha z}\left\{\left|\frac{\sin Tz}{Tz}\right|\left[LD\cos(\beta-\tau)+LB\sin(\beta-\tau)\right]z \pm\left|\frac{\sin\left(\frac{1}{2}Tz\right)}{\frac{1}{2}Tz}\right|^2\left[LB'.CD-LD'.CB\right]\frac{z^2}{2}\right\},$$

$$\left.\begin{matrix} M_{02} \\ M_{20} \end{matrix}\right\} = -e^{-\alpha z}\left\{\left|\frac{\sin Tz}{Tz}\right|\left[LD'\cos(\beta-\tau)+LB'\sin(\beta-\tau)\right]z \pm\left|\frac{\sin\left(\frac{1}{2}Tz\right)}{\frac{1}{2}Tz}\right|^2\left[CB.LD-CD.LB\right]\frac{z^2}{2}\right\},$$

$$\left.\begin{matrix} M_{03} \\ M_{30} \end{matrix}\right\} = e^{-\alpha z}\left\{\left|\frac{\sin Tz}{Tz}\right|\left[CD\cos(\beta-\tau)+CB\sin(\beta-\tau)\right]z \pm\left|\frac{\sin\left(\frac{1}{2}Tz\right)}{\frac{1}{2}Tz}\right|^2\left[LB.LD'-LD.LB'\right]\frac{z^2}{2}\right\},$$

$$\left.\begin{matrix} M_{12} \\ M_{21} \end{matrix}\right\} = e^{-\alpha z}\left\{\pm\left|\frac{\sin Tz}{Tz}\right|\left[CB\cos(\beta-\tau)-CD\sin(\beta-\tau)\right]z +\left|\frac{\sin\left(\frac{1}{2}Tz\right)}{\frac{1}{2}Tz}\right|^2\left[LB.LB'+LD.LD'\right]\frac{z^2}{2}\right\},$$

$$\left.\begin{matrix} M_{13} \\ M_{31} \end{matrix}\right\} = e^{-\alpha z}\left\{\pm\left|\frac{\sin Tz}{Tz}\right|\left[LB'\cos(\beta-\tau)-LD'\sin(\beta-\tau)\right]z -\left|\frac{\sin\left(\frac{1}{2}Tz\right)}{\frac{1}{2}Tz}\right|^2\left[CB.LB+CD.LD\right]\frac{z^2}{2}\right\},$$

$$\left.\begin{matrix} M_{23} \\ M_{32} \end{matrix}\right\} = e^{-\alpha z}\left\{\pm\left|\frac{\sin Tz}{Tz}\right|\left[LD\sin(\beta-\tau)-LB\cos(\beta-\tau)\right]z -\left|\frac{\sin\left(\frac{1}{2}Tz\right)}{\frac{1}{2}Tz}\right|^2\left[LB'.CB+LD'.CD\right]\frac{z^2}{2}\right\}.$$

$$\mathbf{C} = 2e^{-\alpha z}\begin{pmatrix} \left|\cos\left(\frac{1}{2}Tz\right)\right|^2 & \frac{iL^*}{T^*}\sin\left(\frac{1}{2}T^*z\right) & \frac{iL'^*}{T^*}\sin\left(\frac{1}{2}T^*z\right) & -\frac{iC^*}{T^*}\sin\left(\frac{1}{2}T^*z\right) \\ -\frac{iL}{T}\sin\left(\frac{1}{2}Tz\right) & \frac{|L|^2}{|T|^2}\left|\sin\left(\frac{1}{2}Tz\right)\right|^2 & \frac{LL'^*}{|T|^2}\left|\sin\left(\frac{1}{2}Tz\right)\right|^2 & -\frac{LC^*}{|T|^2}\left|\sin\left(\frac{1}{2}Tz\right)\right|^2 \\ -\frac{iL'}{T}\sin\left(\frac{1}{2}Tz\right) & \frac{L'L^*}{|T|^2}\left|\sin\left(\frac{1}{2}Tz\right)\right|^2 & \frac{|L'|^2}{|T|^2}\left|\sin\left(\frac{1}{2}Tz\right)\right|^2 & -\frac{L'C^*}{|T|^2}\left|\sin\left(\frac{1}{2}Tz\right)\right|^2 \\ \frac{iC}{T}\sin\left(\frac{1}{2}Tz\right) & -\frac{CL^*}{|T|^2}\left|\sin\left(\frac{1}{2}Tz\right)\right|^2 & -\frac{CL'^*}{|T|^2}\left|\sin\left(\frac{1}{2}Tz\right)\right|^2 & \frac{|C|^2}{|T|^2}\left|\sin\left(\frac{1}{2}Tz\right)\right|^2 \end{pmatrix}$$

$$
= 2M_{00}\begin{pmatrix}
\cos^2\frac{\phi}{2} & \sin\frac{\phi}{2}\frac{iL^*}{\left(\mathbf{T}\cdot\mathbf{T}^*\right)^{1/2}}e^{i(\beta-\tau)} & \sin\frac{\phi}{2}\frac{iL'^*}{\left(\mathbf{T}\cdot\mathbf{T}^*\right)^{1/2}}e^{i(\beta-\tau)} & -\sin\frac{\phi}{2}\frac{iC^*}{\left(\mathbf{T}\cdot\mathbf{T}^*\right)^{1/2}}e^{i(\beta-\tau)} \\
-\sin\frac{\phi}{2}\frac{iL}{\left(\mathbf{T}\cdot\mathbf{T}^*\right)^{1/2}}e^{-i(\beta-\tau)} & \sin^2\frac{\phi}{2}\frac{\left|L\right|^2}{\left(\mathbf{T}\cdot\mathbf{T}^*\right)^{1/2}} & \sin^2\frac{\phi}{2}\frac{LL'^*}{\left(\mathbf{T}\cdot\mathbf{T}^*\right)^{1/2}} & -\sin^2\frac{\phi}{2}\frac{LC^*}{\left(\mathbf{T}\cdot\mathbf{T}^*\right)^{1/2}} \\
-\sin\frac{\phi}{2}\frac{iL'}{\left(\mathbf{T}\cdot\mathbf{T}^*\right)^{1/2}}e^{-i(\beta-\tau)} & \sin^2\frac{\phi}{2}\frac{L'L^*}{\left(\mathbf{T}\cdot\mathbf{T}^*\right)^{1/2}} & \sin^2\frac{\phi}{2}\frac{\left|L'\right|^2}{\left(\mathbf{T}\cdot\mathbf{T}^*\right)^{1/2}} & -\sin^2\frac{\phi}{2}\frac{L'C^*}{\left(\mathbf{T}\cdot\mathbf{T}^*\right)^{1/2}} \\
\sin\frac{\phi}{2}\frac{iC}{\left(\mathbf{T}\cdot\mathbf{T}^*\right)^{1/2}}e^{-i(\beta-\tau)} & -\sin^2\frac{\phi}{2}\frac{CL^*}{\left(\mathbf{T}\cdot\mathbf{T}^*\right)^{1/2}} & -\sin^2\frac{\phi}{2}\frac{CL'^*}{\left(\mathbf{T}\cdot\mathbf{T}^*\right)^{1/2}} & \sin^2\frac{\phi}{2}\frac{\left|C\right|^2}{\left(\mathbf{T}\cdot\mathbf{T}^*\right)^{1/2}}
\end{pmatrix}.
$$

$$
\mathbf{F} = \sqrt{2}e^{-\alpha z/2}\begin{pmatrix}
\cos\left(\tfrac{1}{2}Tz\right) & 0 & 0 & 0 \\
-\frac{iL}{T}\sin\left(\tfrac{1}{2}Tz\right) & 0 & 0 & 0 \\
-\frac{iL'}{T}\sin\left(\tfrac{1}{2}Tz\right) & 0 & 0 & 0 \\
\frac{iC}{T}\sin\left(\tfrac{1}{2}Tz\right) & 0 & 0 & 0
\end{pmatrix} = \sqrt{2M_{00}}\begin{pmatrix}
\cos\frac{\phi}{2}e^{i\kappa} & 0 & 0 & 0 \\
-\sin\frac{\phi}{2}\frac{iL}{\left(\mathbf{T}\cdot\mathbf{T}^*\right)^{1/2}}e^{i(\kappa-\beta+\tau)} & 0 & 0 & 0 \\
-\sin\frac{\phi}{2}\frac{iL'}{\left(\mathbf{T}\cdot\mathbf{T}^*\right)^{1/2}}e^{i(\kappa-\beta+\tau)} & 0 & 0 & 0 \\
\sin\frac{\phi}{2}\frac{iC}{\left(\mathbf{T}\cdot\mathbf{T}^*\right)^{1/2}}e^{i(\kappa-\beta+\tau)} & 0 & 0 & 0
\end{pmatrix}.
$$

## Appendix

We review the basic definitions and terminology [1, 9]. The electric field vector is $\mathbf{e}$ (complex). The polarization matrix (sometimes called the coherency or coherence matrix) is

$$\mathbf{P} = \left\langle \mathbf{e}\mathbf{e}^{\dagger} \right\rangle, \tag{1}$$

where $\langle\ \rangle$ denotes an average (time or space) and the dagger means complex conjugate transpose. The Stokes matrix $\mathbf{S}$ is the polarization matrix represented in the Pauli basis. Written as a column vector it becomes the Stokes vector $\mathbf{s}$. The Mueller matrix describes the transformation of the Stokes vector from $\mathbf{s}$ to $\mathbf{s}'$:

$$\mathbf{s}' = \mathbf{M}\mathbf{s}. \tag{2}$$

In a similar way, the Mueller matrix in the Cartesian (standard) representation, $\mathbf{N}$, describes the transformation of the polarization vector $\mathbf{p}$ (the polarization matrix written in column vector form) to $\mathbf{p}'$ [10-12]:

$$\mathbf{p}' = \mathbf{N}\mathbf{p}. \tag{3}$$

We take the Pauli matrices in their usual order in optics, but also orthonormalize them ( $\{\boldsymbol{\sigma}_{\mu}, \boldsymbol{\sigma}_{\nu}\} = \mathrm{Tr}\{\boldsymbol{\sigma}_{\mu}^{\dagger}\boldsymbol{\sigma}_{\nu}\} = \delta_{\mu\nu}$ ) [10, 12]:

$$\begin{gathered}\boldsymbol{\sigma}_{\mu} = \frac{1}{\sqrt{2}}\begin{pmatrix} 1 & 0 \\ 0 & 1 \end{pmatrix}, \frac{1}{\sqrt{2}}\begin{pmatrix} 1 & 0 \\ 0 & -1 \end{pmatrix}, \\ \frac{1}{\sqrt{2}}\begin{pmatrix} 0 & 1 \\ 1 & 0 \end{pmatrix}, \frac{1}{\sqrt{2}}\begin{pmatrix} 0 & -i \\ i & 0 \end{pmatrix}.\end{gathered} \tag{4}$$

Then the $4 \times 4$ matrix formed from the four Pauli vectors $\hat{\boldsymbol{\sigma}}_{\mu}$ is the unitary matrix $\boldsymbol{\Lambda}$ (so $\boldsymbol{\Lambda}\boldsymbol{\Lambda}^{\dagger} = \mathbf{I}_4$, the $4 \times 4$ unit matrix) given by [10]

$$\boldsymbol{\Lambda} = \begin{pmatrix} \hat{\boldsymbol{\sigma}}_0 & \hat{\boldsymbol{\sigma}}_1 & \hat{\boldsymbol{\sigma}}_2 & \hat{\boldsymbol{\sigma}}_3 \end{pmatrix} = \frac{1}{\sqrt{2}}\begin{pmatrix} 1 & 1 & 0 & 0 \\ 0 & 0 & 1 & -i \\ 0 & 0 & 1 & i \\ 1 & -1 & 0 & 0 \end{pmatrix}. \tag{5}$$

The Mueller matrix $\mathbf{M}$ represents transformation of the Stokes vector $\mathbf{s}$ to $\mathbf{s}'$ : $\mathbf{s}' = \mathbf{M}\mathbf{s}$, where the 16 elements of the Mueller matrix are real. An Hermitian matrix $\mathbf{C}$, called the coherency matrix, can be generated from the Mueller matrix by [13, 14]

$$\mathbf{C} = \sum_{\mu,\nu} M_{\mu\nu}\boldsymbol{\Gamma}_{\mu\nu} \tag{6}$$

where the $16 \times (4 \times 4)$ (i.e. sixteen $4 \times 4$ ) unitary coherency basis matrices $\boldsymbol{\Gamma}_{\mu\nu}$ are defined by

$$\boldsymbol{\Gamma}_{\mu\nu} = \boldsymbol{\Lambda}^{\dagger}(\boldsymbol{\sigma}_{\mu} \otimes \boldsymbol{\sigma}_{\nu}^{*})\boldsymbol{\Lambda}. \tag{7}$$

Here $\otimes$ means Kronecker (outer) product [13]. We have the relationships $\boldsymbol{\Gamma}_{\mu\nu}^{\dagger} = \boldsymbol{\Gamma}_{\mu\nu}$ ($\boldsymbol{\Gamma}_{\mu\nu}$ is Hermitian), and $\boldsymbol{\Gamma}_{\nu\mu} = \boldsymbol{\Gamma}_{\mu\nu}^{*} = \boldsymbol{\Gamma}_{\mu\nu}^{\mathrm{T}}$, where $^{\dagger}$ signifies conjugate transpose, $^{T}$ signifies transpose and $^{*}$ signifies complex conjugate.

Then

$$\mathbf{p} = \boldsymbol{\Lambda}\mathbf{s}. \tag{8}$$

If the Pauli matrices are trace-normalized, $\{\boldsymbol{\sigma}_{\mu}, \boldsymbol{\sigma}_{\nu}\} = \mathrm{Tr}\{\boldsymbol{\sigma}_{\mu}^{\dagger}\boldsymbol{\sigma}_{\nu}\} = \delta_{\mu\nu}$, then $\boldsymbol{\Lambda}$ is unitary. Then

$$\mathbf{N} = \boldsymbol{\Lambda}\mathbf{M}\boldsymbol{\Lambda}^{\dagger}, \mathbf{M} = \boldsymbol{\Lambda}^{\dagger}\mathbf{N}\boldsymbol{\Lambda}. \tag{9}$$

For a deterministic system, (i.e. one that can be described by a Jones matrix), the electric field vector is transformed from $\mathbf{e}$ to $\mathbf{e}'$ by the Jones matrix $\mathbf{J}$, with column vector $\mathbf{j}$. Then

$$\mathbf{e}' = \mathbf{J}\mathbf{e}. \tag{10}$$

The Jones vector written in the Pauli representation $\mathbf{c}$ is related to $\mathbf{j}$ by

$$\mathbf{j} = \boldsymbol{\Lambda}\mathbf{c}. \tag{11}$$

Then

$$\mathbf{N} = \mathbf{J} \otimes \mathbf{J}^*, \tag{12}$$

so that

$$\mathbf{M} = \mathbf{\Lambda}^\dagger (\mathbf{J} \otimes \mathbf{J}^*) \mathbf{\Lambda}. \tag{13}$$

We define

$$\mathbf{K} = [\mathbf{\Lambda} \otimes (\mathbf{\Lambda}^{-1})^T] \tag{14}$$

where $\otimes$ means Kronecker (outer) product. If $\mathbf{M}$ and $\mathbf{N}$ are written as column vectors $\mathbf{m}, \mathbf{n}$, then [15]

$$\mathbf{n} = \mathbf{K}\mathbf{m}. \tag{15}$$

If $\mathbf{\Lambda}$ is unitary, $\mathbf{\Lambda}\mathbf{\Lambda}^\dagger = \mathbf{I}_4$, the $4 \times 4$ identity matrix, and

$$\mathbf{K} = (\mathbf{\Lambda} \otimes \mathbf{\Lambda}^*) \tag{16}$$

is also unitary, so $\mathbf{m} = \mathbf{K}^\dagger \mathbf{n}$.

$\mathbf{N}$ can be converted into an Hermitian matrix $\mathbf{H}$ by a partial rearrangement of rows operation [11, 12]. $\mathbf{H}$ is the covariance matrix, also called the standard Hermitian matrix. In terms of column vectors this can be written

$$\mathbf{h} = \mathbf{R}\mathbf{n}. \tag{17}$$

The matrix $\mathbf{R}$ is both orthogonal and unitary, so that $\mathbf{R}^{-1} = \mathbf{R}$, and so $\mathbf{n} = \mathbf{R}\mathbf{h}$. $\mathbf{R}$ can be written

$$\mathbf{R} = \mathbf{I}_2 \otimes \mathbf{A}_2 \otimes \mathbf{I}_2, \tag{18}$$

where $(\mathbf{A}_2)^2 = \mathbf{I}_4$, so that also $\mathbf{A}_2^{-1} = \mathbf{A}_2$. It is found that [16]

$$\mathbf{A}_2 = \sum_{\mu=0}^{3} \left( \boldsymbol{\sigma}_\mu \otimes \boldsymbol{\sigma}_\mu^\dagger \right). \tag{19}$$

i.e. equal to the trace of the matrix of the basis matrices $\boldsymbol{\sigma}_\mu \otimes \boldsymbol{\sigma}_\nu^\dagger$.

From Eqs. 15 and 17, we have

$$\mathbf{h} = \mathbf{R}\mathbf{K}\mathbf{m} = \mathbf{\Psi}\mathbf{m}, \tag{20}$$

where

$$\mathbf{\Psi} = \mathbf{R}\mathbf{K}, \tag{21}$$

is unitary. For a deterministic system,

$$\mathbf{H} = \mathbf{j}\mathbf{j}^\dagger = \mathbf{j} \otimes \mathbf{j}^*. \tag{22}$$

This can also be written

$$\mathbf{H} = \mathbf{\Lambda}(\mathbf{c}\mathbf{c}^\dagger)\mathbf{\Lambda}^\dagger = \mathbf{\Lambda}(\mathbf{c} \otimes \mathbf{c}^*)\mathbf{\Lambda}^\dagger. \tag{23}$$

The coherency matrix $\mathbf{C}$, also called the Pauli Hermitian matrix, is given by

$$\mathbf{C} = \mathbf{\Lambda}^\dagger \mathbf{H} \mathbf{\Lambda}, \quad \mathbf{H} = \mathbf{\Lambda}\mathbf{C}\mathbf{\Lambda}^\dagger, \tag{24}$$

[12-14, 17]. For a deterministic system

$$\mathbf{C} = \mathbf{c}\mathbf{c}^\dagger = \mathbf{c} \otimes \mathbf{c}^*. \tag{25}$$

We see that Eqs. 23 and 25 are consistent with Eq. 24.

Also, the similarity of Eqs. 9 and 24, allows us to write in column vector form, using Eq. 15,

$$\mathbf{h} = \mathbf{K}\mathbf{g}, \tag{26}$$

where $\mathbf{H}$ written in column vector form is $\mathbf{h}$, and $\mathbf{C}$ written in column vector form is $\mathbf{g}$. From Eqs. 20 and 26, we then have

$$\mathbf{g} = \mathbf{K}^\dagger \mathbf{h} = \mathbf{\Gamma}\mathbf{m}, \tag{27}$$

where

$$\mathbf{\Gamma} = \mathbf{\Psi}^\dagger \mathbf{R} \mathbf{\Psi} = \mathbf{K}^\dagger \mathbf{R} \mathbf{K}, \tag{28}$$

and $\mathbf{\Psi}\mathbf{\Gamma}\mathbf{\Psi}^\dagger = \mathbf{\Psi}\mathbf{\Gamma}^\dagger\mathbf{\Psi}^\dagger$, so $\mathbf{\Gamma} = \mathbf{\Gamma}^{-1}$ is an involutory matrix.

Finally, we have the relationship

$$\mathbf{g} = \boldsymbol{\Psi}^\dagger \mathbf{n} = \mathbf{K}^\dagger \mathbf{R}\mathbf{n}. \tag{29}$$

Hence the matrices, written in column vector form, can be calculated from each other using vector multiplications by $\mathbf{R}$ and $\mathbf{K}$.

The polarization matrix is transformed according to

$$\mathbf{P}' = \mathbf{J}\mathbf{P}\mathbf{J}^\dagger. \tag{30}$$

An arbitrary, non-deterministic polarization matrix can be decomposed by summing over the four eigenvalues (called the Kraus decomposition [18], like a coherent mode decomposition)

$$\begin{aligned} \mathbf{P}' &= \sum_{\alpha=0}^{3} \mathbf{J}_\alpha \mathbf{P} \mathbf{J}_\alpha^\dagger, \\ \mathbf{J}_\alpha &= \sqrt{\lambda_\alpha}\,\bar{\mathbf{J}}_\alpha, \end{aligned} \tag{31}$$

where $\bar{\mathbf{J}}_\alpha$ is a normalized Jones matrix. This allows a combination of Jones matrices to be used to represent a non-deterministic system. A similar form for the Stokes matrix, i.e. the Stokes vector written in matrix form, rather than the Jones matrix, seems not to exist [19].

For a deterministic system, the coherency matrix factorizes, $\mathbf{C} = \mathbf{c}\mathbf{c}^\dagger$ (Eq. 25). The coherency matrix is non-negative definite, and any non-negative definite Hermitian matrix can be factorized into the product of a matrix and its conjugate transpose. So we can factorize any coherency matrix, whether deterministic or not, into the product of a matrix $\mathbf{F}$ and its conjugate transpose [20]. We have

$$\begin{aligned} \mathbf{C} &= \mathbf{U}\mathbf{D}\mathbf{U}^\dagger = \sum_{\alpha=0}^{3} \lambda_\alpha \bar{\mathbf{C}}_\alpha = \sum_{\alpha=0}^{3} \mathbf{U}\mathbf{D}_\alpha \mathbf{U}^\dagger = \sum_{\alpha=0}^{3} \left(\mathbf{U}\sqrt{\mathbf{D}_\alpha}\right)\left(\mathbf{U}\sqrt{\mathbf{D}_\alpha}\right)^\dagger = \sum_{\alpha=0}^{3} \mathbf{F}_\alpha \mathbf{F}_\alpha^\dagger = \mathbf{F}\mathbf{F}^\dagger, \\ \bar{\mathbf{C}}_\alpha &= \mathbf{u}_\alpha \mathbf{u}_\alpha^\dagger, \mathbf{F}_\alpha = \mathbf{U}\sqrt{\mathbf{D}_\alpha}, \\ \mathbf{F} &= \mathbf{U}\,\mathrm{diag}\left(\sqrt{\lambda_0}, \sqrt{\lambda_1}, \sqrt{\lambda_2}, \sqrt{\lambda_3}\right), \end{aligned} \tag{32}$$

where $\lambda_\alpha$ are the eigenvalues of $\mathbf{C}$, $\mathbf{D}$ is the diagonal matrix of eigenvalues, $\mathbf{D}_\alpha$ is the matrix of a single eigenvalue, $\mathbf{u}_\alpha$ are the normalized eigenvectors of $\mathbf{C}$, and $\mathbf{U}$ the matrix of eigenvectors. The square root of a diagonal matrix is equal to the matrix of the square roots of the elements. We call $\mathbf{F}$ the coherency matrix factor (CMF). It consists of the four component c-vectors, weighted by the square roots of the coherency matrix eigenvalues, and arranged in column vector form.

In the same way that we can factorize the coherency matrix, we can factorize the covariance matrix $\mathbf{H}$, to give the covariance matrix factor (the matrix of weighted Jones-vectors), $\mathbf{E}$ [20]:

$$\begin{aligned} \mathbf{H} &= \mathbf{V}\mathbf{D}\mathbf{V}^\dagger = \sum_{\alpha=0}^{3} \lambda_\alpha \bar{\mathbf{H}}_\alpha = \sum_{\alpha=0}^{3} \mathbf{V}\mathbf{D}_\alpha \mathbf{V}^\dagger = \sum_{\alpha=0}^{3} \left(\mathbf{V}\sqrt{\mathbf{D}_\alpha}\right)\left(\mathbf{V}\sqrt{\mathbf{D}_\alpha}\right)^\dagger = \sum_{\alpha=0}^{3} \mathbf{E}_\alpha \mathbf{E}_\alpha^\dagger = \mathbf{E}\mathbf{E}^\dagger, \\ \bar{\mathbf{H}}_\alpha &= \mathbf{v}_\alpha \mathbf{v}_\alpha^\dagger, \mathbf{E}_\alpha = \mathbf{V}\sqrt{\mathbf{D}_\alpha}, \\ \mathbf{E} &= \mathbf{V}\,\mathrm{diag}\left(\sqrt{\lambda_0}, \sqrt{\lambda_1}, \sqrt{\lambda_2}, \sqrt{\lambda_3}\right), \end{aligned} \tag{33}$$

where $\lambda_\alpha$ are the eigenvalues of $\mathbf{H}$, equal to the eigenvalues of $\mathbf{C}$, $\mathbf{v}_\alpha$ are the normalized eigenvectors of $\mathbf{H}$, and $\mathbf{V}$ the matrix of eigenvectors. We call $\mathbf{E}$ the covariance matrix factor. It consists of the four component Jones vectors, weighted by the square roots of the covariance matrix eigenvalues, and arranged in column vector form.